\documentclass[useAMS,usenatbib]{mnras}
\usepackage[]{natbib,amsmath,amssymb,times,refname,bm}
\bibpunct{(}{)}{;}{a}{}{,}
\usepackage{tabularx,amsmath,amssymb,hyperref}
\usepackage{graphicx,epsfig,color,latexsym,nicefrac,textcomp}

\renewcommand{\d}{\mathrm{d}}

\newcommand{\ii}{\mathrm{i}}
\newcommand{\bea}{\begin{eqnarray}}
\newcommand{\eea}{\end{eqnarray}}
\newcommand{\be}{\begin{equation}}
\newcommand{\ee}{\end{equation}}
\newcommand{\rund}[1]{\left(#1\right)}

\defcitealias{Semboloni13}{S13}

\long\def\/*#1*/ {}

\def\elabel#1{\label{eq:#1}}

\title[Colour gradient bias]
{Calibration of colour gradient bias in shear measurement using HST/CANDELS data}
\author[X. Er et al.]%
{
X. Er$^{1,2}$ \thanks{xer@ynu.edu.cn},
H. Hoekstra$^3$, T. Schrabback$^4$, V. F. Cardone$^1$, R. Scaramella$^{1,7}$, R. Maoli$^{5,7}$,
\newauthor{M. Vicinanza$^{1,5,6,8}$, B. Gillis$^{9}$, J. Rhodes$^{10,11}$}
\\
$^1$ I.N.A.F. - Osservatorio Astronomico di Roma, via Frascati 33, 00040 - Monte Porzio Catone, Roma, Italy\\
$^{2}$ South-Western Institute for Astronomy Research, Yunnan University, Kunming, P.R.China\\
$^3$Leiden Observatory, Leiden University, PO Box 9513, NL-230 RA, Leiden, the Netherlands \\
$^4$Argelander Instutite f\"ur Astronomie, Auf dem H\"ugel 71, D-53121 Bonn, Germany\\
$^5$Dipartimento di Fisica, Universita di Roma "La Sapienza", Piazzale Aldo Moro, 00185 - Roma, Italy\\
$^6$Dipartimento di Fisica, Universita di Roma "Tor Vergata", via della Ricerca Scientifica 1, 00133 - Roma, Italy\\
$^7$I.N.F.N. - Universit$\grave{a}$ id Roma "La Sapienza", Ple Aldo Moro 2, -Roma, Italy\\
$^8$Instituto de Astrof\'{i}sica e Ci\^{e}ncias do Espa\c{c}o, Universidade de Lisboa,
Faculdade de Ci\^{e}ncias, Campo Grande, PT1749-016 Lisbon, Portugal\\
$^9$Royal Observatory, University of Edinburgh, Blackford Hill, Edinburgh EH9 3HJ, UK\\
$^{10}$Jet Propulsion Laboratory, California Institute of Technology, 4800 Oak Grove Drive, Pasadena, CA 91109, USA\\
$^{11}$California Institute of Technology, 1200 East California Blvd, Pasadena, CA 91125, USA
}
\date{Accepted --;  received --;  in original from \today}
\pubyear{2018}

\begin{document}
\maketitle

\begin{abstract}
Accurate shape measurements are essential to infer cosmological parameters from large area weak gravitational lensing studies. The compact diffraction-limited point-spread function (PSF) in space-based observations is greatly beneficial, but its chromaticity for a broad band observation can lead to new subtle effects that could hitherto be ignored: the PSF of a galaxy is no longer uniquely defined  and spatial variations in the colours of galaxies result in biases in the inferred lensing signal. Taking {\it Euclid} as a reference, we show that this colour-gradient bias (CG bias) can be quantified with high accuracy using available multi-colour {\it Hubble} Space Telescope (HST) data. In particular we study how noise in the HST observations might impact such measurements and find this to be negligible. We determine the CG bias using HST observations in the F606W and F814W filters and observe a correlation with the colour, in line with expectations, whereas the dependence with redshift is weak. The biases for individual galaxies are generally well below 1\%, which may be reduced further using morphological information from the {\it Euclid} data. Our results demonstrate that CG bias should not be ignored, but it is possible to determine its amplitude with sufficient precision, so that it will not significantly bias the weak lensing measurements using {\it Euclid} data.
\end{abstract}


\section{Introduction}

The images of distant galaxies are distorted, or sheared, by the tidal effect of the gravitational  potential generated by intervening matter; an effect commonly referred to as weak gravitational lensing \citep[see e.g.][for a detailed introduction]{Bartelmann01}. The resulting correlations in the shapes can be related directly to the statistical properties of the mass distribution in the Universe, which in turn depend on cosmological parameters. Hence weak gravitational lensing by large-scale structure, or cosmic shear, has been identified as a powerful tool for cosmology. The measurement of the signal as a function of cosmological time is sensitive to the expansion history and the growth rate of large-scale structures, and thus can be used to constrain models for dark energy and modified gravity \citep{2016arXiv160600180A}.

A useful measurement of the cosmic shear signal requires averaging over large numbers of galaxies
to reduce the uncertainty caused by the intrinsic ellipticities of galaxies. The result is, however, only meaningful if biases in the shape estimates are negligible. Various instrumental effects change the observed ellipticities by more than the typical lensing signal, which is of order one per cent. The most dominant source of bias is the smearing of the images by the point spread function (PSF), driving the desire for space-based observations \citep{Paulin-Henriksson08, Massey13}.
Despite these observational challenges, the most recent cosmic shear studies are starting to yield competitive constraints on cosmological parameters \citep{Heymans13,Jee16,Hildebrandt17,DESSV, 2017arXiv170801538T}. These results are based on surveys of modest areas of the sky, which limits their ability to study the nature of dark energy; to achieve that requires more than an order of magnitude improvement in precision.

Such a measurement is the objective of {\it Euclid}\footnote{www.euclid-ec.org}
\citep{Laureijs11}, the dark energy mission of the European Space
Agency (ESA) that will survey the 15\,000 deg$^2$ of extragalactic sky
with both low extinction and low zodiacal light. To reduce the
detrimental effects of noise on the shape measurements, the images
used for the lensing analysis are observed using a wide bandpass
(550-920\,nm). The much smaller PSF in space-based observations is a
major advantage, but the diffraction-limited PSF leads to new
complications.

The most prominent one is that the correction for the smearing by the chromatic PSF depends on the spectral energy distribution (SED) of the galaxy of interest \citep{Cypriano10, Eriksen17} and ignoring this would lead to significant biases in the case of {\it Euclid}. Fortunately this can be accounted for using the supporting broad-band observations that are used to derive photometric redshifts for the sources: the correction employs an effective PSF that is derived from the estimate of the observed SED of the galaxy. This correction is sufficient if the SED does not vary spatially. If this is not the case, the underlying brightness distribution, which is needed for an unbiased estimate of the shear, cannot be unambiguously recovered from the observed images.
This results in a higher-order systematic bias, which we call colour-gradient bias (or CG bias in short).
As shown by \cite{Semboloni13} (\citetalias{Semboloni13} hereafter) the amplitude depends on several factors: the SED of the galaxy, the relative size of the galaxy compared to the PSF, and the width of the bandpass, $\Delta\lambda$.  For instance, the bias scales as $\Delta\lambda^2$, and thus is particularly relevant in the case of {\it Euclid}.

Galaxies show a wide variety in colour gradients,  caused by differences in the properties
of the underlying stellar populations. In particular, elliptical galaxies (ETGs) show mostly
negative colour gradients (redder in the centre and bluer in the outskirts), with steeper gradients more commonly found in bluer or more luminous ETGs \citep[e.g.][]{2005ApJ...635..243F, 2011MNRAS.414.3052D, 2011MNRAS.411.1151G}. Comparison of these colour gradients with population
synthesis models suggest a dominant radial trend in metallicity for red sequence ETGs
\citep[e.g.][]{2011ApJ...740L..41L,2016A&A...593A..84K}. However, towards $z>0.5$ a sizeable
fraction of ETGs display blue cores, caused by a substantial population of young stars in these
galaxies, a trend that can be expected to increase with redshift \citep{2009MNRAS.395..554F, 2010ApJS..187..374S}. In contrast, the more complex distribution of age and metallicity in
late-type galaxies translates into different dependencies \citep{2005ApJ...630..784T}. Hence the relation between galaxy morphology and density may cause the CG bias to vary across the sky and may lead to correlations with the lensing signal itself.

It is important that all systematic sources of biases are accounted for to a level that is smaller than the
statistical uncertainties. In the case of {\it Euclid} this leads to tight requirements, as detailed in
\cite{Massey13} and \cite{Cropper13}. Initial studies by \cite{Voigt12} and \citetalias{Semboloni13} used simulated images to show that the CG bias could be substantial,
exceeding nominal requirements for the multiplicative bias in the shear. They also argued
that it should be possible to calibrate the bias using {\it Hubble} Space Telescope (HST) observations
of a large sample of galaxies in the F606W and F814W filters. However, their conclusions are based on the analysis of simulated noiseless data. In this work, we revisit the issue of the calibration of CG bias,
with a particular focus on determining the bias from data with realistic noise levels.

In Sect.~\ref{sec:concepts}, we describe the main concepts and introduce the notation. We present the results from the analysis of simulated images in Sect.~\ref{sec:simulations}. In particular we explore the  impact of having to use noisy data to measure the CG bias in Sect.~\ref{sec:noisy}.
In Sect.~\ref{sec:candels} we estimate the CG bias using HST observations from the Cosmic Assembly Near-infrared Deep Extragalactic Legacy Survey \citep[CANDELS;][]{Koekemoer11}.

\section{The origin of colour gradient bias}
\label{sec:concepts}

Following the notation of \citetalias{Semboloni13}, we consider an image of a galaxy, and denote the photon brightness
distribution of the image at each position $\bm \theta$ and wavelength $\lambda$ by $I({\bm \theta};\lambda)$, which is related to the intensity $S({\bm\theta};\lambda)$ by $I^0({\bm\theta};\lambda)=\lambda S({\bm\theta};\lambda)
T(\lambda)$, where $T(\lambda)$ is the normalised transmission. We take this to be a top-hat with a
width $\Delta\lambda$ around a central wavelength $\lambda_{\rm cen}$. The resulting image of the galaxy, observed using a telescope with a PSF $P({\bm \theta};\lambda)$  is given by:
\be
I^{\rm obs}({\bm\theta}) = \int_{\Delta\lambda} I^0({\bm\theta}; \lambda) *
P({\bm \theta},\lambda)\, \d \lambda,
\label{eq:iobs}
\ee
where $*$ denotes a convolution.

A measurement of the ellipticity of a galaxy provides an unbiased (but noisy) estimate of the
weak gravitational lensing signal, quantified by the complex shear $\gamma=\gamma_1+\ii\gamma_2$.
The ellipticity $\epsilon$ in turn can be determined from the second order brightness moments $Q^0_{ij}$ of the PSF-corrected image $I^0(\theta)$:
\be
\epsilon_1+\ii \epsilon_2 \approx \frac{Q^0_{11} - Q^0_{22} + 2 \ii Q^0_{12} }
{Q^0_{11} + Q^0_{22} +2(Q^0_{11}Q^0_{22} - (Q^0_{12})^2)^{1/2}}
\elabel{mshear}
\ee
where the second order brightness moments are given by\footnote{We implicitly assume that the moments are evaluated around the position where the dipole moments vanish.}
\be
Q^0_{ij} = {1 \over F} \int  I^{0}({\bm \theta})\, \theta_i \theta_j \, \d^2 {\bm \theta} \quad\; (i,j=1,2),
\ee
where $F=\int \d^2{{\bm\theta}}  I^{0}({\bm\theta})$ is the total observed photon flux.

\begin{figure*}
\includegraphics[width=13.5cm]{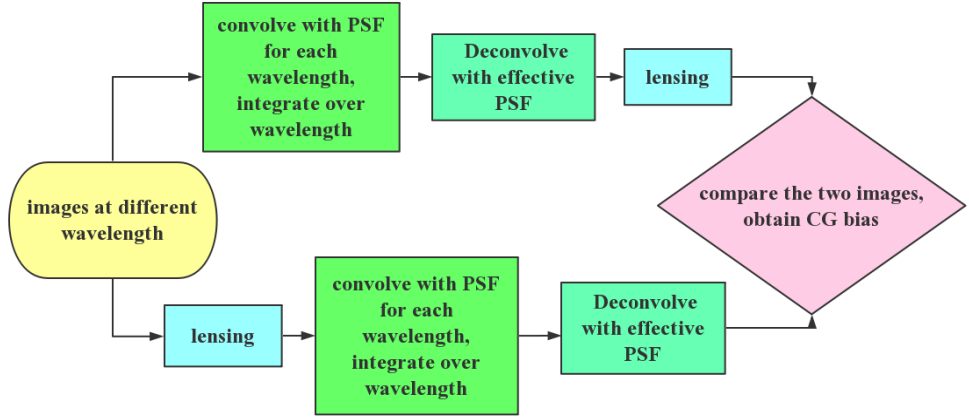}
\caption{Flowchart describing how the colour-gradient bias is determined. The initial image
is the same in both cases, but in the top flow an image without a colour gradient is created
to which a shear is applied. In the bottom flow, the image is sheared before the PSF steps
are applied. The ellipticities of the resulting images differ slightly, and can be used to quantify
the bias that is introduced.}
\label{fig:flowchart}
\end{figure*}

In practice, however, the observed moments are measured from the PSF-convolved image given by
Eqn.~(\ref{eq:iobs}). Moreover, the moments are evaluated using a weight function $W({\bm\theta})$  to reduce the effect of noise in the images. Hence, the observed quadrupole moments are given by
\be
Q^{\rm obs}_{ij} = {1 \over F_{\rm w}} \int_{\Delta\lambda} \d\lambda \int \d^2 {\bm\theta} \,
I^0({\bm\theta}; \lambda) *  P({\bm\theta},\lambda)\, \theta_i \theta_j \, W({\bm \theta})\,,
\ee
where $F_{\rm w}$ is the weighted flux. The use of a weight function biases the observed moments,
and the aim of moment-based shape measurement algorithms is to correct for this using estimates of the higher order moments \citep[e.g.][]{Kaiser1995,Melchior11}. An alternative approach is to fit sheared, PSF-convolved models to the observed images
\citep[e.g.][]{Bridle02,2007MNRAS.382..315M,2008MNRAS.390..149K,Miller13}; in these fitting methods the profile itself acts as a weight.

\citetalias{Semboloni13} showed that the inevitable use of a weight function gives rise to the CG bias.
Consequently, the bias depends on the choice of the weight function, and vanishes in the case of {\it unweighted} moments. In the latter case it is possible to determine the PSF-corrected moments $Q^0_{ij}$ from the observed quadrupole moments because
\be
Q^{\rm obs}_{ij}=Q^0_{ij}+P^{\rm eff}_{ij} \,\label{eq:unweighted}
\ee
for unweighted moments, where $P^{\rm eff}_{ij}$ are the quadrupole moments of the effective PSF,  defined as
\be
P^{\rm eff}({\bm \theta}) = \frac{1}{F} \int \d \lambda\, P({\bm \theta},\lambda)\, F(\lambda) \,,
\ee
where $F(\lambda)$ is the photon flux as a function of wavelength, which is directly related to the spectral energy distribution (SED) of the galaxy. Hence the correction for the chromatic PSF requires an estimate of the SED.  \cite{Eriksen17} have shown that the broadband observations that are used to determine photometric redshifts for {\it Euclid} can also be used to estimate the effective PSF with sufficient accuracy to meet the stringent requirements presented in \cite{Cropper13}.

We limit our study of the CG bias to the multiplicative bias it
introduces, and our approach to quantify the impact on the lensing
signal is similar to \citetalias{Semboloni13}.
Fig.~\ref{fig:flowchart} shows the flowchart of the steps that enable
us to evaluate the CG bias. In both cases we start with the same
wavelength-dependent image $I^0({\bm \theta};\lambda)$, but the bottom
flow resembles what happens in the actual observations: the original
image is sheared\footnote{We use $\gamma_1=0.05$ and $\gamma_2=0.02$
  as reference, but we verified that other values yield similar
  results (difference smaller than $1\%$).} before the convolution
with the PSF. The deconvolution with the effective PSF then yields the
PSF-corrected shape\footnote{We perform the deconvolution of the
  effective PSF in Fourier space (see Eqs.\,(12) and (13) in
  \citetalias{Semboloni13}). For the images with noise, we deconvolve
  the best fit image, i.e. without the residual pixel noise.}.  In the
top flow the PSF steps are applied first, resulting in an image
without a colour gradient that is subsequently sheared.

We measure the ellipticities of the resulting images to estimate the CG bias. To reduce noise in our estimate of the multiplicative bias $m$ we use the ring-test method \citep{Nakajima07} where we create eight copies of the original galaxy (i.e. pre-lensed and pre-PSF convolution) but with different orientations. The ensemble-averaged ellipticities then  provide an estimate of the multiplicative CG bias, $m$ (we do not explore additive bias here), via
\be
m= {\langle \epsilon^{\rm CG}\rangle \over \langle \epsilon^{\rm NCG}\rangle }-1,
\ee
where `CG' indicates the case where the galaxy has a colour gradient, and `NCG' is the galaxy
with a uniform colour. Note that our approach differs slightly from that in \citetalias{Semboloni13}, who quantify the response of the observed ellipticity to an applied shear. Consequently their definition
of $m$ has the opposite sign. The procedural difference with \citetalias{Semboloni13} is that they do not apply the last step in the bottom flow (the deconvolution), but rather convolve the final image in the top flow.
The steps presented in Fig.~\ref{fig:flowchart} yield a more symmetric result, highlighting the fact
that the CG bias is the consequence of the fact that the shearing of the image does not commute
with the convolution with the PSF. However, we verify in Sect.~\ref{sec:simulations} that we recover
the results of \citetalias{Semboloni13} (but with an opposite sign).

Recently, \cite{Huff17} proposed a technique to infer multiplicative shear calibration parameters that avoids the use of extensive image simulations, such as those described in \citep{Hoekstra17}. They quantify the
sensitivity to a known shear by applying it to the observed data. Hence, their approach follows the top flow in Fig.~\ref{fig:flowchart} and thus cannot account for CG bias.

\section{Colour gradient bias in simulated data}
\label{sec:simulations}

The CG bias is a higher-order systematic bias, and thus the changes in the measured ellipticities are small. It is therefore important to verify that numerical errors in the calculations are subdominant compared to the small effects we aim to measure. To do so, we compare results from two independent codes that are used to generate the simulated images: one is written in C/C++ and the other uses the
python-based {\sc GalSim} package \citep{Rowe15}, which is widely used to created simulated images \citep[e.g.][]{FenechConti17, Hoekstra17,2017arXiv170801533Z}.

In the C/C++ code we compute the images using a sheared S{\'e}rsic
profile, and multiply the surface brightness at the centre of each
pixel with the pixel area. In the case of {\sc GalSim} we use the {\sc
  Shear()} function (which convolves the image by the pixel). Since we
are interested in small differences in the shapes of deconvolved
images, we first examined the size of potential numerical errors. We
therefore convolved and subsequently deconvolved elliptical
images. Comparison of the recovered ellipticities revealed small
multiplicative differences between the codes that ranged from
$10^{-7}$ to $10^{-6}$, two orders of magnitude smaller than the CG
biases we are concerned with. Hence can safely neglect these numerical
artefacts here.

As a further test we compare directly to the results obtained by \citetalias{Semboloni13} for two reference galaxy models. The main purpose is to validate the approach of measuring the CG bias using noisy images and estimating the necessary number of galaxies to perform the calibration for the {\it Euclid} lensing survey. Although the choice of the galaxy models cannot represent the full sample of {\it Euclid}, the resulting colour gradients are sufficiently large so that the CG bias will not be underestimated. The small galaxy model represents the smallest galaxies that will be used in the {\it Euclid} weak lensing analysis \citep{Massey13}. Therefore the current two galaxy models are sufficient for this aspect of our work.
The reference galaxies are modeled as the sum of a bulge and disk component. To describe the wavelength dependence of the images we use the galaxy SED templates from \citet{1980ApJS...43..393C}: we use the SED for an elliptical galaxy for the bulge and take the SED of an irregular galaxy for the disk. The two components are  described by a circular S{\'e}rsic profile:
\be
I_{\rm S}(\theta) = I_0 {\rm e}^{-\kappa\; \left(\frac{\theta}{r_{\rm h}}\right)^{1/n}},
\ee
where $I_0$ is the central intensity, and $\kappa=1.9992\,n -0.3271$. For the bulge component we adopt $n=1.5$ and for the disk we use $n=1$. The profiles are normalised such that the bulge contains 25\% of the flux at a wavelength of 550\,nm. The galaxies are circular and the half-light radii, $r_{\rm h}$, for the bulge and disk for galaxy `B'   are $0\farcs17$ and $1\farcs2$, respectively. The second galaxy `S' is smaller with half-light radii of $0\farcs09$ and $0\farcs6$  for the bulge and disk, respectively (also see Table~3 in \citetalias{Semboloni13}). We create images with a size of  $256\times256$ pixels, and resolution $0.05$ arcsec/pixel at wavelengths 1\,nm apart and sum these in the range $550-920$\,nm to mimic the {\it Euclid} pass-band.

\begin{figure}
\centerline{\includegraphics[width=\hsize]{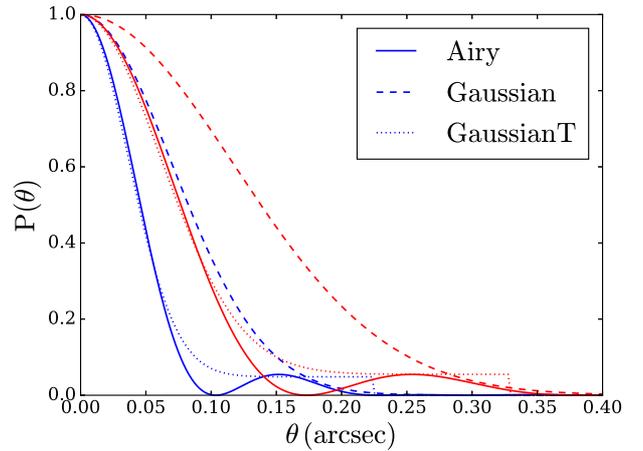}}
\caption{Comparison of the obscured Airy profile (solid), which is a good approximation
to the {\it Euclid} PSF, to PSF1 (Gaussian; dashed) and PSF3 (compact Gaussian and
top-hat; dotted) from \citetalias{Semboloni13}. The profiles for 550\,nm are indicated by the blue lines and
the results for 920\,nm are shown in red.}
\label{fig:psfmodel}
\end{figure}

To create the PSF-convolved images we consider several PSF profiles. For a direct comparison with \citetalias{Semboloni13} we use their reference PSF1. As discussed in \citetalias{Semboloni13} this PSF has a similar size as the nominal {\it Euclid} PSF, but a steeper wavelength-dependence. Our implementation of the pipeline was able to reproduce the results presented in \citetalias{Semboloni13}.
To better approximate the {\it Euclid} PSF \citetalias{Semboloni13} also considered a model that  consists of a compact Gaussian core and an appropriately scaled top-hat (their PSF3). Instead we use here a more realistic obscured Airy profile, which is actually close to the {\it Euclid} design profile \citep{Laureijs11}:
\be
P(x) = {I_0 \over (1-\epsilon^2)^2} \rund{{2J_1(x)\over x} -
{2\epsilon J_1(\epsilon x) \over x}}^2,
\elabel{psfairy}
\ee
where $I_0$ is the maximum intensity at the centre, $\epsilon$ is the aperture obscuration ratio, and $J_1(x)$ is the first kind of Bessel function of order one; $x$ is defined as $x=\pi \theta/(\lambda\, D) $.
In the case of {\it Euclid}, $D=1.2$m and $\epsilon=1/3$. We compare this model to the Gaussian
case and PSF3 from \citetalias{Semboloni13} in Fig.~\ref{fig:psfmodel} at 550\,nm and 920\,nm.

\begin{figure}
  \centerline{\includegraphics[width=\hsize]{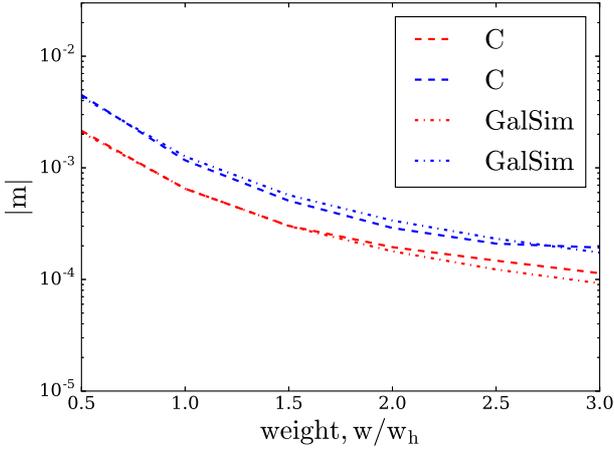}}
\caption{The CG bias versus width of the weight function (in
  units of the half-light radius $w_{\rm h}$) used to compute the
  quadrupole moments for the large (`B'; red) and small (`S';
  blue) reference galaxy. The galaxies were convolved using the obscured Airy
  PSF. The dashed (dash-dotted) lines are our
  results for images simulated using the {\sc C} ({\sc GalSim}) code.}
\label{fig:biasofweight}
\end{figure}

As discussed in Sect.~\ref{sec:concepts} the amplitude of the bias depends on the width of the weight function that is used to compute the (weighted) quadrupole moments. In Fig.~\ref{fig:biasofweight} we show the CG bias for the two reference galaxies as a function of $\theta_{\rm w}$, the width of the weight function that is used to compute the quadrupole moments. The results from the {\sc C} code (dashed lines) and the {\sc GalSim} code (dotted lines) agree very well for both the large galaxy `B' (red lines) and the small galaxy `S' (blue lines). Given the consistent results between the {\sc C} and {\sc GalSim} code we conclude that numerical errors are negligible in our implementation. In the remainder, we limit the simulations to those generated with {\sc GalSim}.

Fig.~\ref{fig:biasofweight} shows that the CG bias decreases rapidly when the width of the weight function is increased. This allows for an interesting trade-off between CG bias and noise bias. The latter increases with increasing $\theta_{\rm w}$ but relatively slowly (see Fig.~4 in \citetalias{Semboloni13}). It also highlights that the CG bias itself differs between shape measurement methods, which typically use different weight functions. As a proxy for the optimal weight function (which maximizes the signal-to-noise ratio) we adopt the value of the  half-light radius in the remainder of this paper. This yields $m=0.65\times10^{-3}$ for galaxy `B' and $m=1.17\times10^{-3}$ for galaxy `S', demonstrating that the CG bias is a strong function of galaxy size.

\subsection{Impact in high-density regions}

The focus of this paper is to quantify the impact of CG bias on cosmic shear measurements,
i.e. we consider only small distortions in the shapes of the sources. However, {\it Euclid} will
also enable the calibration of the masses of galaxy clusters with unprecedented precision.
\cite{Koehlinger15} have shown that this should be possible given the accuracy required
for the shape measurement algorithms for cosmic shear. This does implicitly assume that the
performance does not change in high density environments. Blending does impact the performance
\citep{Hoekstra17}, but can be accounted for. In this section we focus instead on the unexplored question
whether the CG bias differs in the central regions of galaxy clusters.
\begin{figure}
  \centerline{\includegraphics[width=\hsize]{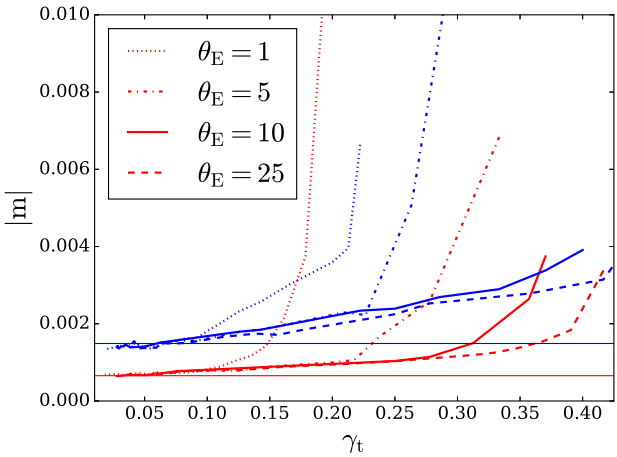}}
  \caption{The CG bias versus tangential shear when the full lens
    equation is used to compute the image distortions. The red lines
    indicate the resulting CG bias for the `B' galaxy, whereas the
    blue lines correspond to the `S' galaxy. The horizontal lines
    indicate the CG bias when we only use shear in the image
    distortions. The bias depends on the Einstein radius, $\theta_{\rm
      E}$, of the lens, and is more prominent for small values of
    $\theta_{\rm E}$ at a given shear amplitude.}
  \label{fig:biasofgamma}
\end{figure}
In high density regions, higher order distortions of the images can
become dominant. For instance, flexion (the next order after shearing)
has been studied as a potential observational tool
\citep[e.g.][]{2002ApJ...564...65G,bacon2006,2011MNRAS.412.2665V}. Rather
than simply shearing the images, as we have done so far, in this
section we use the full lens equation to perform ray tracing
simulations instead. This enables us to capture the effect of the
higher order distortion. For this exercise we use the {\sc C} code, as
it has this functionality fully implemented. As a lens we consider a
singular isothermal sphere (SIS) with an Einstein radius $\theta_{\rm
  E}$; in this case the (tangential) shear is given by $\gamma_{\rm
  t}(\theta)=\;$\textonehalf$\,\theta_{\rm E}/\theta$. To minimise numerical
effects, the image sizes are increased to $2048\times2048$ pixels,
with a resolution $0\farcs0125$/pixel.

In Fig.~\ref{fig:biasofgamma} we show the CG bias as a function of the tangential shear
for different values of $\theta_{\rm E}$. The red lines indicate the results for the `B' galaxy
and the blue lines show the biases for the `S' galaxy. For small shears, i.e. far away from the lens,
the CG bias converges to the shear-only case that we have studied thus far (the thin horizontal lines).
Hence, for cosmic shear studies we can safely ignore this complication. However, as the source approaches the lens, the flexion signal increases, resulting in an increase in the  CG bias. The change depends on the value of $\theta_{\rm E}$, because flexion is lower for a given shear when the source is further away from the lens. Hence, the additional CG bias due to higher order distortions is expected to be relatively small for clusters of galaxies (for which $\theta_{\rm E}>10''$), but it can be relevant for studies
of massive galaxies; in this case the Einstein radius is smaller, and the flexion signal larger. Fig.~\ref{fig:biasofgamma} shows that for a lens with $\theta_{\rm E}=1''$ the CG bias rapidly increases when the shear $\gamma>0.15$, i.e. for $\theta<3''$. Thanks to the small PSF of {\it Euclid} it is possible to measure the galaxy-galaxy lensing signal on such small scales, which could in principle provide interesting constraints on the enclosed stellar mass. However, our findings indicate that colour gradients may complicate the measurement of the small-scale galaxy-galaxy lensing signal. This warrants further study that is beyond the scope of this paper.

\begin{figure*}
  \centerline{
  \includegraphics[width=8.5cm]{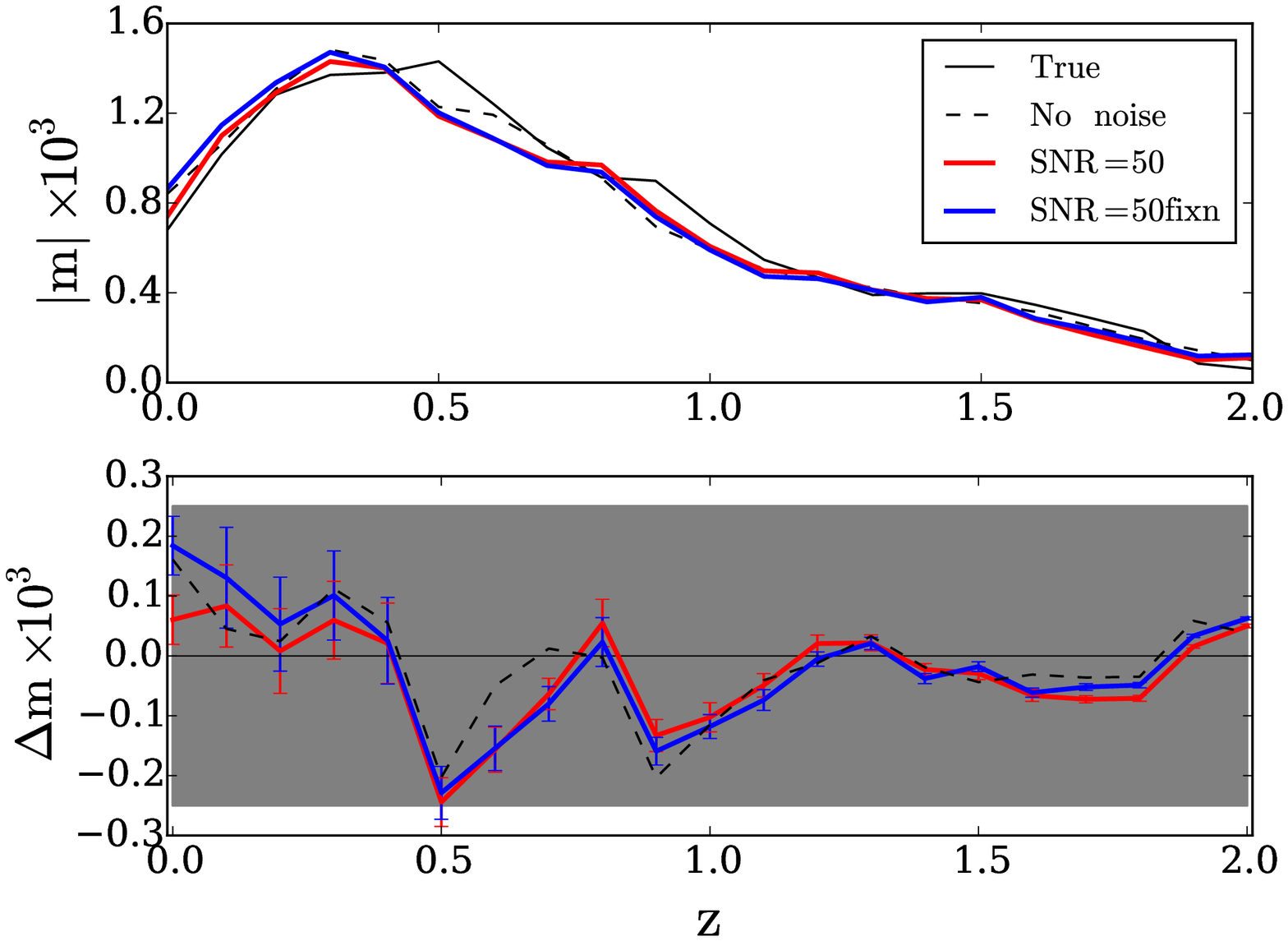}
  \includegraphics[width=8.5cm]{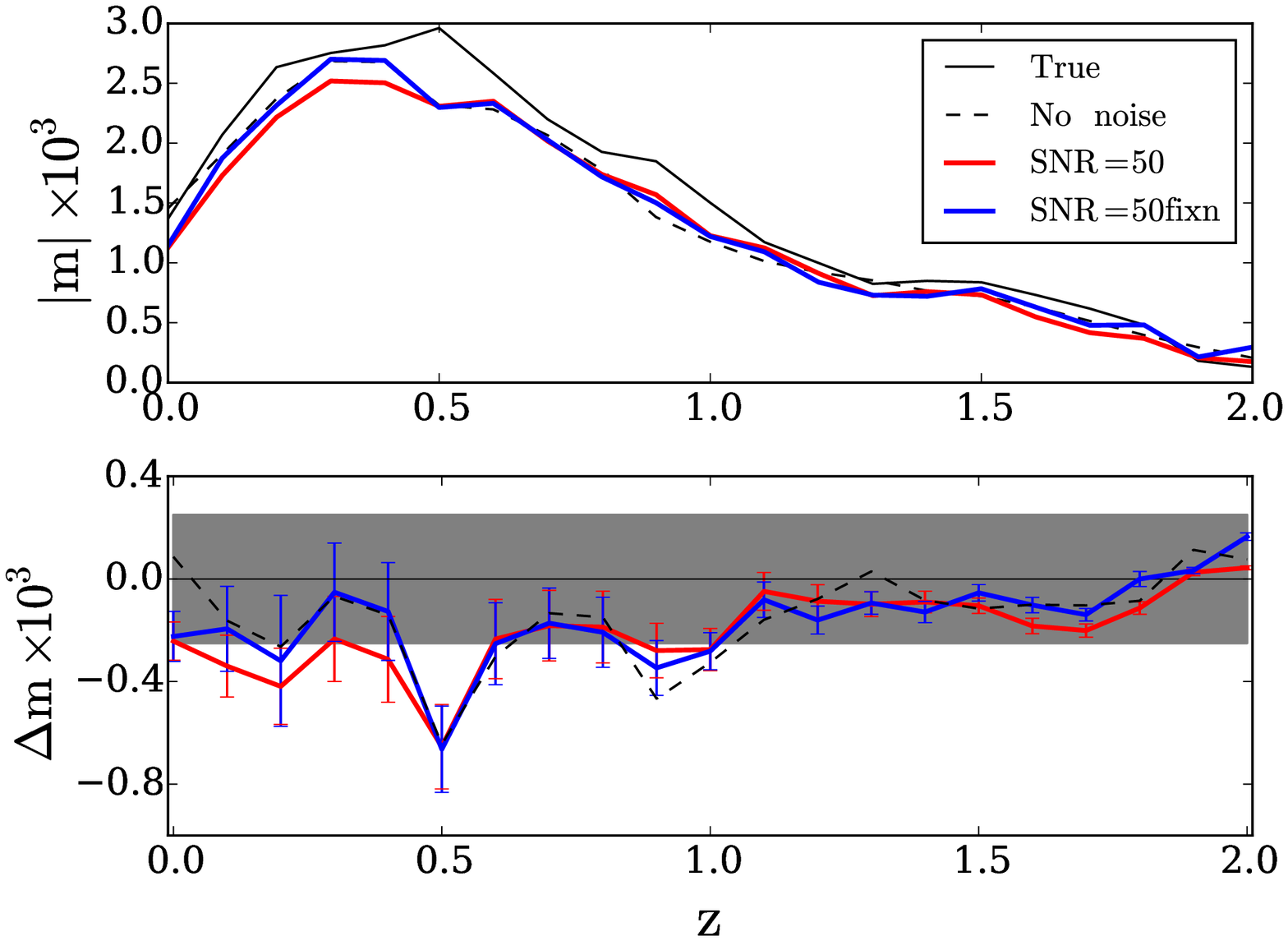}}
\caption{The multiplicative CG bias as a function of redshift
  for the reference galaxies, with the results for galaxy `B' shown
  in the left panel and those for galaxy `S' in the right panel.
  The dashed black line is the recovered bias when we mimic
  noiseless HST observations in two filters. The solid red line
  indicates the results when we use the best fit {\sc {\tt GALFIT}} model
  in both filters to estimate the CG bias when the simulated HST images have
  an input $\mathrm{SNR}=50$ (averaged over 40 noise realisations at each
  redshift). The blue line shows the results when we fix the S{\'e}rsic index
  in the fit. The bottom panels show the residuals $\Delta m$ with respect
  to the true CG bias. The grey band indicates the
  nominal {\it Euclid} requirement for the residual CG bias after correction.}
\label{fig:biasofz50}
\end{figure*}

\subsection{Calibration of CG bias using simulated HST images}
\label{sec:noisy}

The {\it Euclid} observations lack high-resolution multi-band images to measure the CG bias directly for each source galaxy. However, the cosmological lensing signal is typically inferred from the ellipticity correlation function, which involves  averaging the shapes of large ensembles of galaxies.
Provided the average bias that is caused by colour gradients is known for a selection of sources, it is possible in principle to obtain unbiased estimates of the ellipticity correlation function.
Here it is particularly important that the correction for the CG bias accounts for the variation in redshift and colour.
The former is relevant for tomographic cosmic shear studies, whereas the latter avoids significant spatial variation in the bias because of the correlation between galaxy colour, or morphology, and density.

\citetalias{Semboloni13} showed that HST observations in both the F606W and F814W filters can be used to determine the CG bias to meet {\it Euclid} requirements. However, \citetalias{Semboloni13} did not consider the complicating factor that the HST images themselves are noisy. Although the HST observations are typically deeper than the nominal {\it Euclid} data, and the HST PSF is considerably smaller, it is nonetheless necessary to investigate the impact of noise in more detail. We address this particular question here, before we determine the CG bias from actual HST data in Sect.~\ref{sec:candels}.
Since we linearly interpolate the SED using two HST bands, the colour gradient within the two bands themselves is lost. Usually this approximation will not cause large deviations in the estimate of the bias, but it may fail when there are strong emission lines (as we will see in the following). The impact of emission lines, their prevalence, etc., requires further analysis that is beyond the scope of this paper.

The method to calibrate the CG bias using observations in two bands is described in detail in \citetalias{Semboloni13}, but here we outline the main steps for completeness. To model the wavelength dependence of the image we use two narrow-band\footnote{To distinguish these filters from the broad VIS pass-band (which has $\lambda$ transmission in the wavelength range $550$nm to $920$nm), we refer to the F606W and F814W as narrow bands, but acknowledge that these are commonly referred to broad-band filters and that genuine narrow-band filters are significantly narrower. The adopted wavelength range for the transmission of the F606W filter is $470$nm$<\lambda<719$nm and $680$nm$<\lambda<960$nm for the F814W filter. } images, each of which is given by:
\be
I_i({\bm\theta}) = \int_{\Delta \lambda_i} T_i(\lambda)\, I({\bm \theta},\lambda) \;\d \lambda,
\elabel{linearitp}
\ee
where $T_i(\lambda)$ is the transmission of the $i$th narrow filter. We assume that for each pixel the wavelength dependence of the image can be interpolated linearly:
\be
I({\bm \theta},\lambda) \approx  a_0({\bm \theta})+a_1({\bm \theta})\lambda.
\elabel{interpolate}
\ee
Eqs.\ref{eq:linearitp} and \ref{eq:interpolate} yield a linear set of
equations for each pixel, which can be used to solve for the
coefficients $a_i$:
\be
T_{0i} a_0(\theta) \,+\,T_{1i} a_1(\theta) = I_i(\theta), \quad\; i=1,2,
\elabel{lineareq}
\ee
where we defined
\be
T_{ji}=\int_{\Delta\lambda_i} \d \lambda\,T_i(\lambda)\lambda^j.
\ee
We thus obtain approximate galaxy images at each wavelength, which we use to estimate the CG bias, following the same procedure as we used in the previous section.

We first consider the recovery of the CG bias for noiseless observations of the two reference galaxies, as this represents the best-case scenario. We simulate the images in the F606W and F814W filters at different redshifts. We adopt the native sampling of the Advanced Camera for Surveys (ACS) on HST of  $0\farcs05$ pixel$^{-1}$. As shown in \citetalias{Semboloni13}, we cannot ignore the blurring of the observed images by the HST PSF; to mimic this we assume an obscured Airy function for a mirror with diameter $D=2.5$m and obscuration  $0.33$ as a proxy for the HST PSF. We deconvolve our synthetic HST images and create the images at different wavelengths as the starting point for the flow presented in Fig.~\ref{fig:flowchart}.
{\bf }

Following \citetalias{Semboloni13}, we show  the CG bias as a function of redshift for galaxy `B' (left panels) and `S' (right panels) in Fig.~\ref{fig:biasofz50}, demonstrating that the CG bias varies significantly with redshift. Note that we ignored any evolution in the galaxy SEDs, which will occur in practice. The results for the actual CG bias are indicated by the solid black lines, whereas the dashed black lines indicate the recovered values from the noiseless synthetic HST observations in the F606W and F814W filters.  The bottom panels show the residuals between the recovered and the true bias. The residual bias is within the target tolerance for {\it Euclid}, indicated by the grey band, for all redshifts.

We now proceed to explore the impact of noise in the HST images. To do so, we add Gaussian noise to the simulated HST images, where the r.m.s. noise level $\sigma$ is determined by the signal-to-noise ratio of the galaxy, SNR; the total flux within an aperture of radius $1.5\times r_{\rm h}$, $F_{\rm tot}$; and the number of pixels within this aperture, $N_{\rm tot}$, such that
\be
\sigma = {F_{tot} \over {\rm SNR} \sqrt{N_{tot}} }.
\ee
For reference, we compared the input SNR for the two reference
galaxies to that estimated by {\sc SExtractor} \citep{Bertin96}
(e.g. we use {\sc FLUX\_AUTO} in the estimation).  We find good
agreement for galaxy `B' for SNR values ranging from 5 to 50 in both
HST filters. The agreement is also good for the `S' galaxy, but {\sc
  SExtractor} returns lower values if the input SNR is larger than 30.
We consider two noise levels for the simulated HST data, which correspond
to a \mbox{$\mathrm{SNR}=50$} and \mbox{$\mathrm{SNR}=15$}. For simulated
HST data with a depth matching the real HST data analysed in
Sect.~\ref{sec:candels}, \mbox{$\mathrm{SNR}=15$} corresponds to
magnitudes $m_{606}=25.7$ and $m_{814}=25.3$ in the HST bands
or approximately a VIS magnitude of $m_{\rm VIS}=25.4$. This is
significantly fainter than the galaxies included in the {\it Euclid}
weak lensing analysis. For comparison, \mbox{$\mathrm{SNR}=50$}
corresponds roughly to $m_{\rm VIS} = 23.7$, a bit brighter than
the typical galaxy used in the {\it Euclid} weak lensing analysis.
\begin{figure*}
  \includegraphics[width=8.5cm]{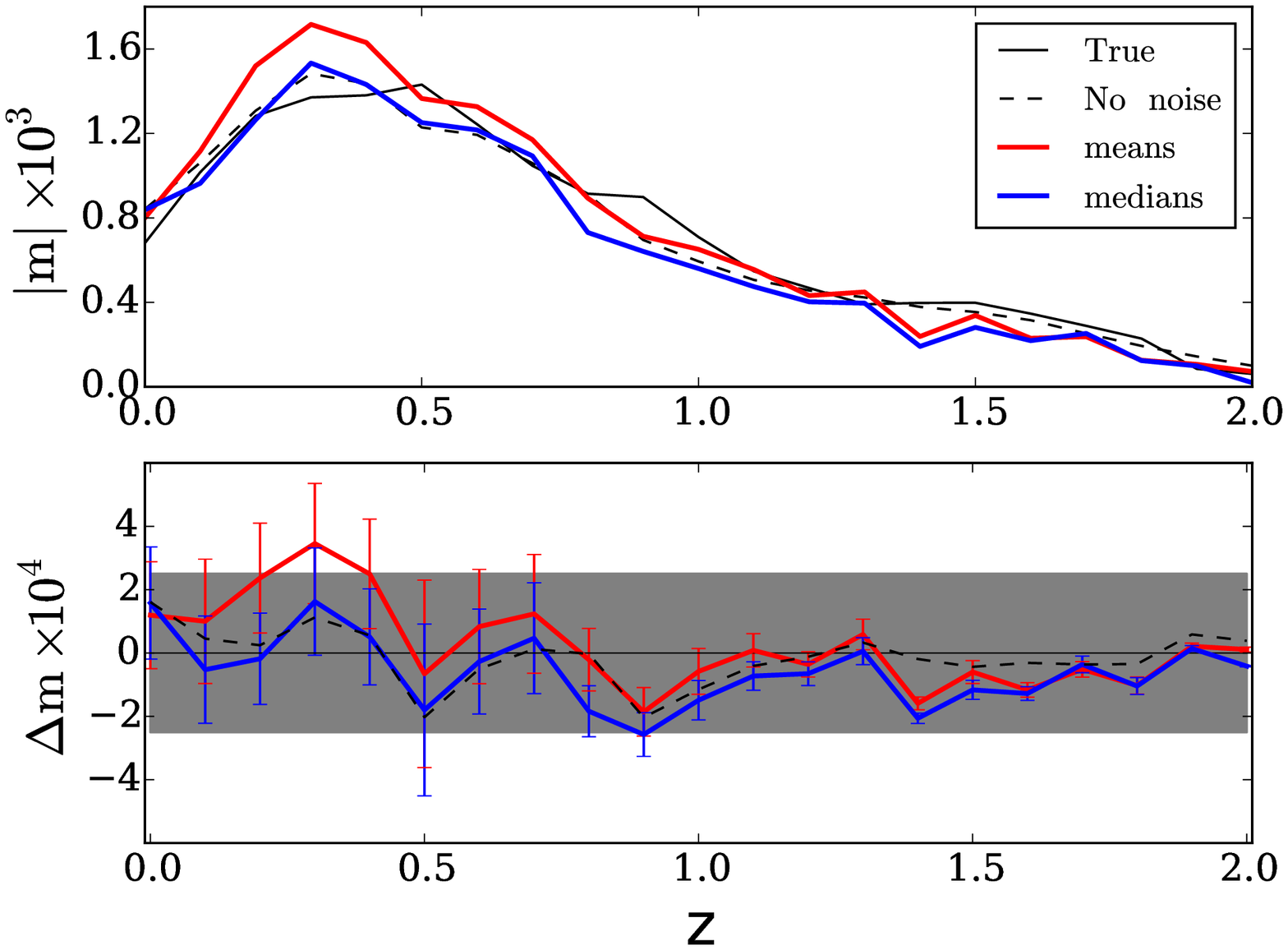}
  \includegraphics[width=8.5cm]{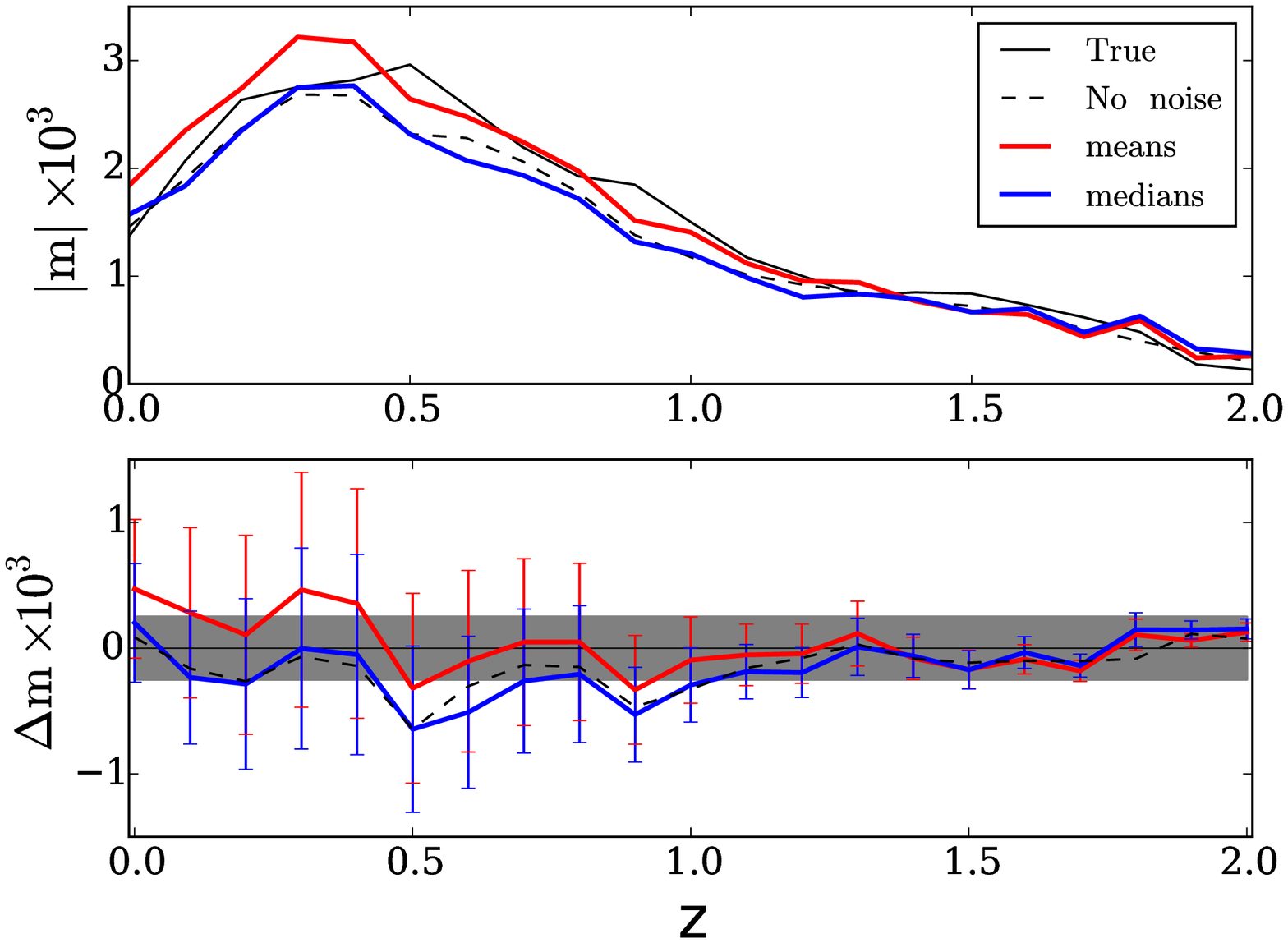}
\caption{Same as Fig.\ref{fig:biasofz50} but for images of
  $\mathrm{SNR}=15$. The red curve shows the mean bias, whereas the
  blue curve corresponds to the median.  To compute the error-bars,
  $1000$ realizations are used at each redshift bin for both the B- and
  the S-galaxy.}
\label{fig:biasofz15}
\end{figure*}
The deconvolution of noisy images is problematic, because the presence of noise will lead to biased estimates of the underlying galaxy. Instead we regulate the problem by assuming that galaxies can be fit by a bulge and disk component, each described by a S{\'e}rsic profile. Real galaxies have more complex morphologies, including spiral arms, etc. To leading order, however, the radial surface brightness profile is the most important quantity, because we are interested in ensemble averages of large numbers of sources with random position angles: morphological features tend to average out in this case. As an additional test we also fitted the galaxies with a single S{\'e}rsic profile. Although the results differ from the true CG bias, depending on e.g. the SNR and morphology of the galaxy, the main trend with redshift is recovered. Nonetheless, further investigation with realistic morphologies is needed, but we conclude that our approach should capture the main properties of the CG bias in real data.

We fit the bulge and disk model, convolved with the PSF, to  the noisy images in each band and use the best fit model to compute the CG bias. To perform the fit, we use {\sc {\tt GALFIT}} \citep{Peng10} with
the prior constraints on the galaxy parameters (S{\'e}rsic index, effective radius, and axis ratio) listed in Table~\ref{fitpar}. We combine the images in the two filters and use {\sc SExtractor} to estimate the centre and some of the initial galaxy parameters to be used as the starting point by {\sc {\tt GALFIT}}. The resulting best-fit images depend somewhat on these initial values, and thus could affect the  estimate for the CG bias. This will be more important when the SNR of the images is lower. To explore this we perform the fits using two sets of initial parameters: in the first we leave all parameters free, while in the other case we fix the S{\'e}rsic index to its simulated value, but leave the other parameters free.
\begin{table}
\begin{center}
\begin{tabular}{ccc|cc}
\hline\hline
parameter &S-606W  & S-814W  & B-606W & B-814W \\ \hline
$n_1$ &0.5--2.5  &0.5--2.5  &0.5--2.5 &0.5--2.5 \\ \
$n_2$ &0.5--2.5  &0.5--2.5  &0.5--2.5 &0.5--2.5 \\ \
$R_{\rm bulge}$ &1--10 &1--10  &3--30  &3--30  \\
$R_{\rm disk}$  &5--30 &5--30  &10--60 &10--60 \\
$q$      &0.6--1  &0.6--1  &0.6--1 &0.6--1 \\
\hline
\end{tabular}
\caption{\label{fitpar} Constraints for the fitting parameters in {\sc {\tt GALFIT}}.
The first two columns are for two images of the S-galaxy, the other two are
the image of B-galaxy.
$n_1$ is the S{\'e}rsic index for bulge, and $n_2$ is the S{\'e}rsic index for disk.
The effect radius is given in unit of pixel ($0.05$ arcsec).}
\end{center}
\end{table}
\begin{figure}
  \includegraphics[width=8.0cm]{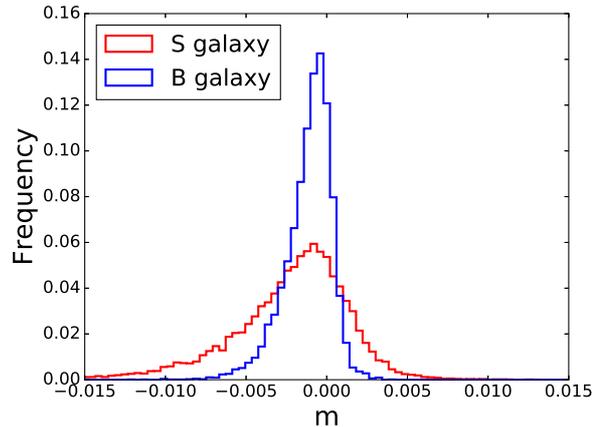}
  \caption{Histogram of the individual noisy estimates of inferred CG
    bias for the `B' (blue) and `S' (red) galaxy when narrow band
    observations with $\mathrm{SNR}=15$ are used. The histogram
    combines the results for the different redshifts. For comparison,
    Fig.\,\ref{fig:biasofz15} shows the mean and median of the noisy
    estimates as a function of redshifts.}
  \label{fig:histogrambias}
\end{figure}

We use the best fit models to compute the CG bias, following the
algorithm that was used to compute the signal in the noiseless
case. We show the resulting average inferred CG bias in
Fig.~\ref{fig:biasofz50} for \mbox{$\mathrm{SNR}=50$} as a function of
redshift for the two reference galaxies (`B' in the left panel and `S'
in the right panel). The bottom panels in Fig.~\ref{fig:biasofz50}
show the residuals $\Delta m$ with respect to the true multiplicative
CG bias. To determine the average bias we analyse six rotations of the
galaxy and use the average value as our estimate of the galaxy
ellipticity \citep{Nakajima07}.  Moreover we create 40 noise
realisations for each redshift to estimate the statistical uncertainty
in our estimate of the multiplicative CG bias, which is  simply a
combination of the uncertainties of images with and without colour
gradient, and is given by
\be
\sigma_m= |m| \sqrt{\rund{\sigma_{\rm cg} \langle e_{\rm cg} \rangle \over \langle e_{\rm ncg}\rangle^2 }^2
  + \rund{\sigma_{\rm ncg} \over \langle e_{\rm ncg} \rangle}^2 },
\elabel{sigmam}
\ee
where $\sigma_{\rm ncg}$ and $\sigma_{\rm cg}$ are the uncertainties in the average
ellipticities for the images without and with a colour gradient, respectively.

We find that fixing the S{\'e}rsic index (blue line) or leaving all parameters free (red line) results in a similar CG bias as a function of redshift. Moreover, the results closely resemble the noiseless case (dashed lines).
The residuals presented in the bottom panel of Fig.~\ref{fig:biasofz50} show that  for the \mbox{$\mathrm{SNR}=50$} case, we expect that the average CG bias can be determined with an overall accuracy that meets the adopted {\it Euclid} tolerance, indicated by the grey band. Only for the `S' galaxy is the residual outside the nominal range at low redshifts, but we note that the reference galaxies have rather extreme colour gradients.
Moreover, the significant deviations at $z=0.5$ and $0.9$ arise because the adopted SED of the disk (Irr) contains strong emission lines  \citepalias[see Fig.~1 in][]{Semboloni13}. These lines enter and exit the F606W filter at these redshifts, respectively, and the linear approximation for the wavelength
dependence fails. In these, albeit extreme cases, two-band imaging may not be sufficient. To what extent this will affect the estimate of the CG bias requires further study.

Fig.~\ref{fig:biasofz15} shows the mean and median of the inferred CG bias for galaxies `S' and `B' as a function of redshift when estimated from noisier simulated images with \mbox{$\mathrm{SNR}=15$}. As for the case with \mbox{$\mathrm{SNR}=50$}, the bias is recovered to a level that is acceptable for {\it Euclid}. Note that we did increase the number of noise realisation to $1000$ to ensure robust estimates of the average CG bias. As expected, the CG bias estimates from the individual noisy images have a larger scatter with a slightly skewed distribution. In Fig.~\ref{fig:histogrambias}, we show the distribution of the CG bias combining results for the full redshift range (\mbox{$\mathrm{SNR}=15$}). Given this increased scatter, a larger sample of real HST galaxy images will be required at these \mbox{$\mathrm{SNR}$} levels in order to calibrate the CG bias at sufficient precision (see Sect.~\ref{sec:candels}).


\begin{figure}
\includegraphics[width=\hsize]{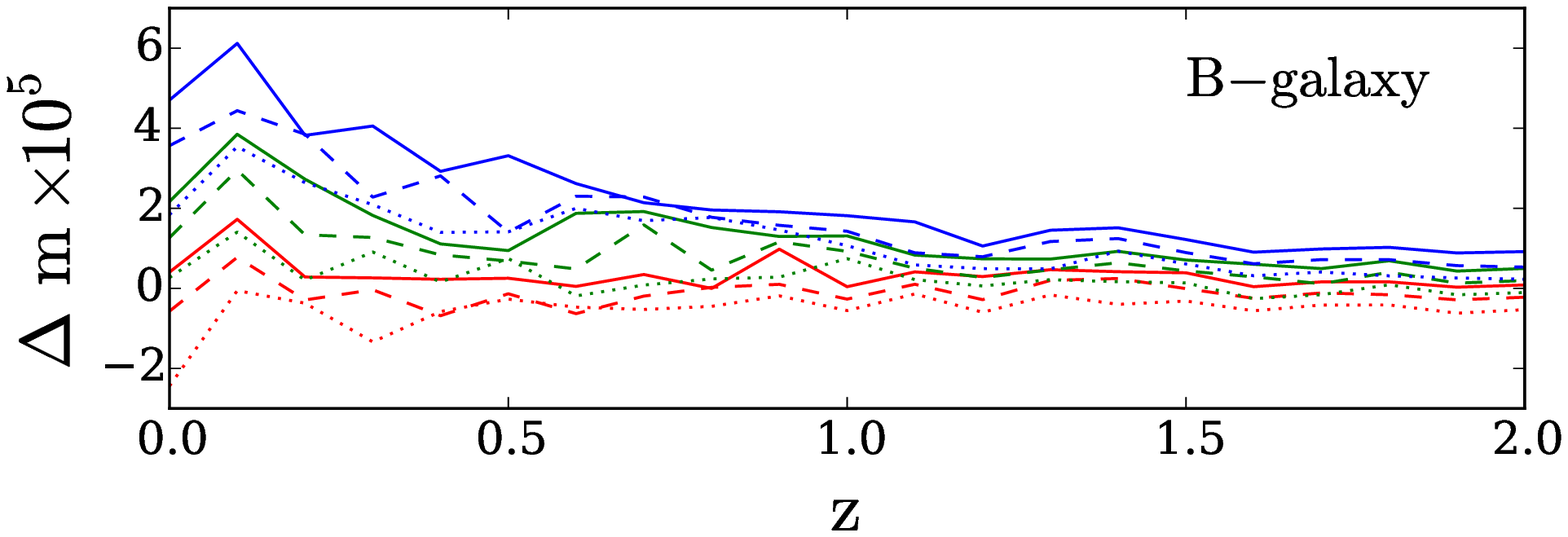}
\includegraphics[width=\hsize]{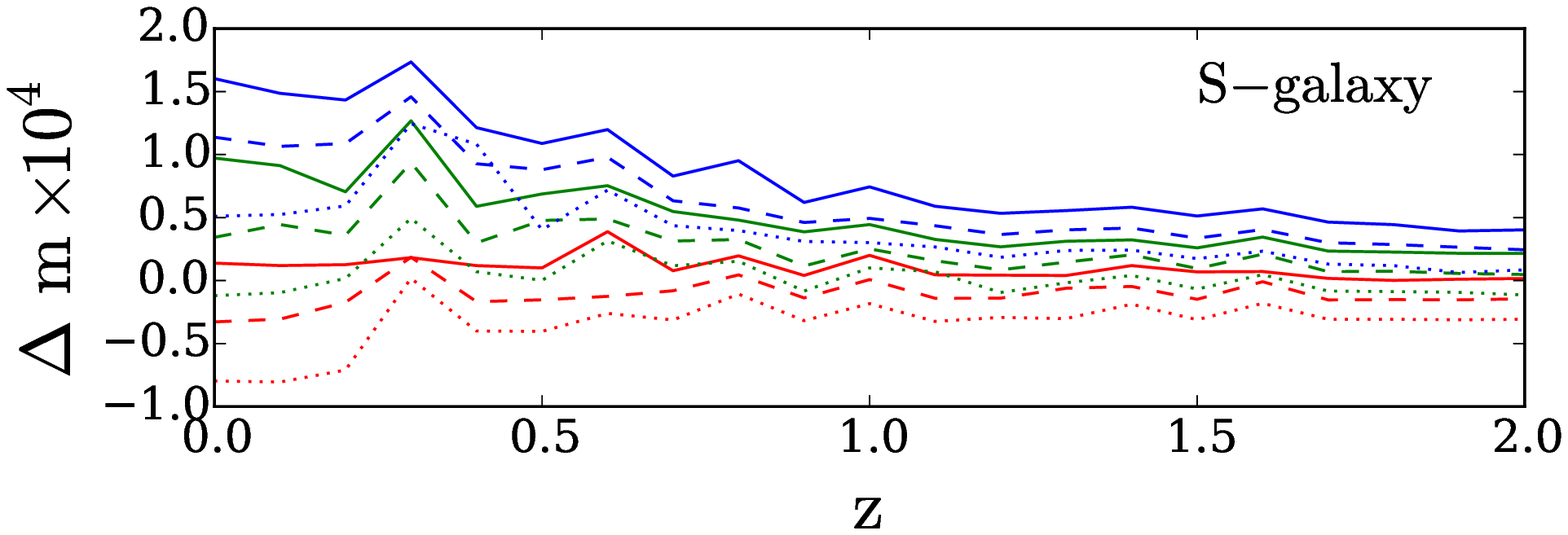}
\caption{Change in multiplicative CG bias when the size of the PSF used in the deconvolution
of the narrow band images is increased (the FWHM differs by 5\% between steps). From red,
green to blue lines, we increase the size of the PSF for the F814W filter; from the solid, dashed to dotted
lines we  increase the size of the PSF for the F606W images.}
\label{fig:psfacc1}
\end{figure}

\subsection{PSF variations in narrow-band data}
\label{sec:psfmodel}

So far we implicitly assumed that the simple axisymmetric PSF used to mimic the HST data
is perfectly known. In reality, however, the HST PSF is more complex, and varies spatially
and as a function of time. The small field-of-view of ACS typically results in a relatively small number
of stars that can be used to model the PSF, although most of the variation can be captured with
few parameters \citep[e.g.][]{Schrabback10}; of these focus variations are the most dominant. We therefore examine next how well the HST PSF  properties need to be determined so that they do not affect the CG bias measurement significantly.

To do so, we first generate models where we slightly increase the PSF size in the two bands by computing the  Airy profile when the wavelength in the calculation is increased by a factor 1.05, 1.10 and 1.15 for the three cases. This increases the effective PSF FWHM in three steps of 5\% between the different cases. These models are used only in the step where we deconvolve the simulated HST images in the absence of noise. The change in CG bias, $\Delta m$ as a function of redshift is shown in Fig.~\ref{fig:psfacc1} for the `B' galaxy (top panel) and `S' galaxy (bottom panel).  The results for an increase in the PSF size in  the F606W band are indicated by the solid, dashed and dotted lines, respectively; the red, green and blue lines indicated the impact of increasing the size of the PSF in the F814W band. The sensitivity to the PSF errors is typically larger for low-redshift galaxies, but the change in CG bias is much smaller than the bias itself.  As expected, small galaxies are more sensitive to errors in the estimate of the PSF size.

\begin{figure}
\includegraphics[width=8.0cm]{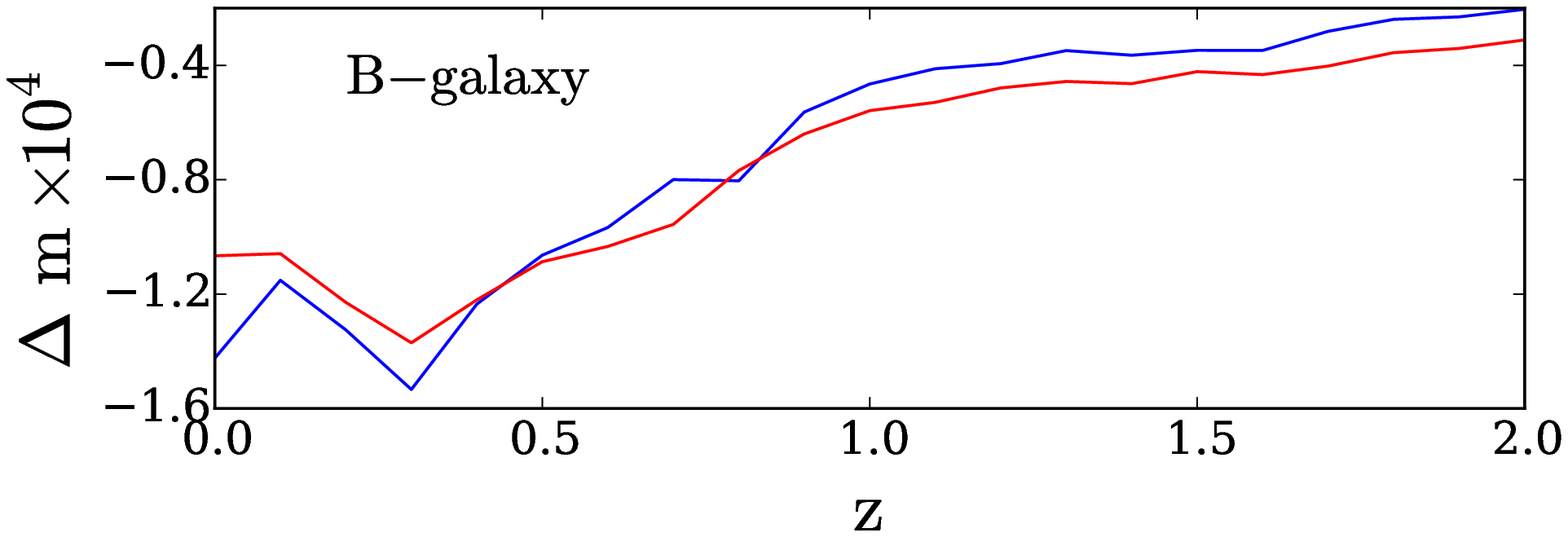}
\includegraphics[width=8.0cm]{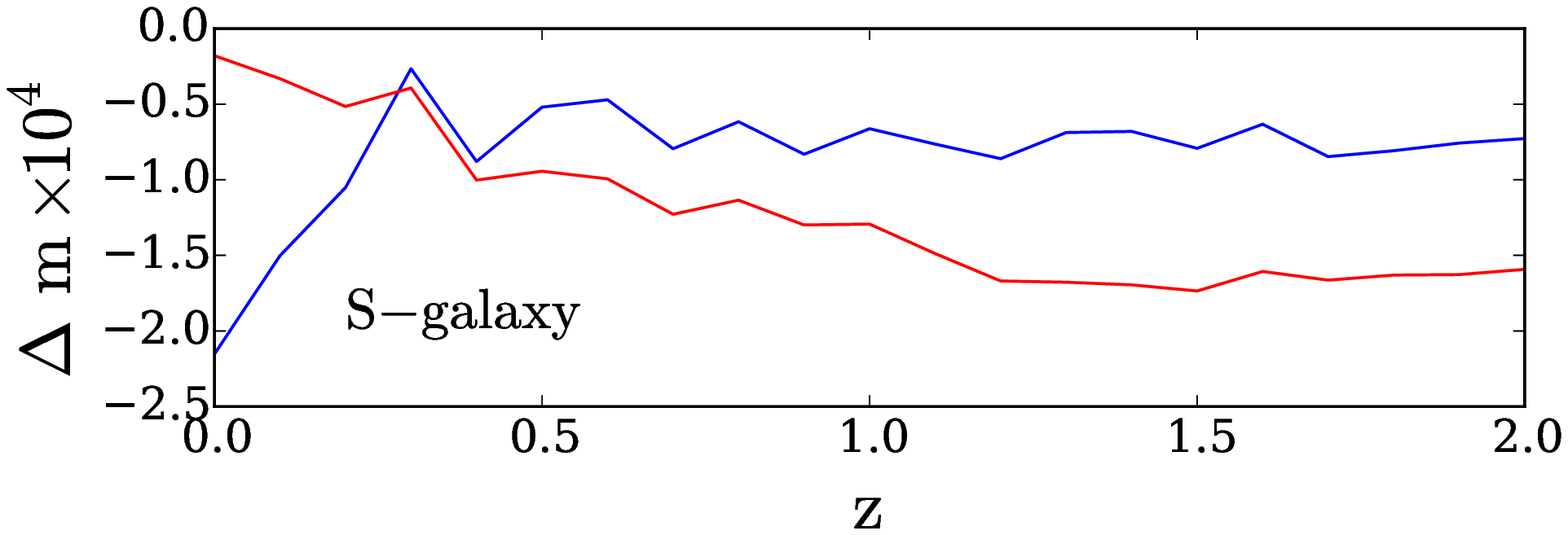}
\caption{Difference in CG bias when the reference {\sc TinyTim} PSF is used to deconvolve
the synthetic HST data (blue lines) or when we mimic the PSF modelling (red lines). The top
panel shows the results for the reference galaxy `B', whereas the bottom panel shows results
for galaxy `S'. The differences are small, suggesting that the bias is not particularly sensitive to
errors in the adopted HST PSF model.}
\label{fig:psfacc2}
\end{figure}

To mimic a more realistic scenario we generated mock star fields using simulated PSFs generated with
the {\sc TinyTim} tool \citep{2011SPIE.8127E..0JK}. To compute the reference PSFs at the various positions on the detector in the F606W and F814W filters we used the default parameters where possible, including the appropriate camera, detector, and filter passband settings for each image. We adopt the K7V spectrum for the SED, which represents a typical stellar SED in the sample (the choice of a fixed spectrum for stars was found to have a negligible impact on the models.). We select stars with a signal-to-noise ratio larger than $50$, and ensure there are no detected objects within $1$ arcsecond ($20$ pixels), and outlier rejection is performed based on the measured moments and sizes of the stars. The postage stamps of the star images for each filter are normalised and then stacked using inverse-variance weighting. The FWHM is $30\%$ larger than the Airy model in the simulation. This PSF is then used to determine the colour gradient bias from the synthetic HST images of the two reference galaxies (which are convolved with an obscured Airy function for a mirror with diameter $D = 2.5$m and obscuration 0.33 as a proxy for the HST PSF). The blue lines in Fig.~\ref{fig:psfacc2} show the resulting difference in CG bias for the `B' (top panel) and `S' galaxy (bottom panel) as a function of redshift. Although this represents a rather significant mismatch in PSF, the change in bias is quite small.

To mimic modeling errors that would occur in reality we select simulated PSF images at a nearby position on the detector (from a grid of points) and fitted for the focus values (for details, see Gillis et al. in prep.). The corresponding model PSFs are stacked using the same weights as before. The resulting change in CG bias for the `B' (top panel) and `S' galaxy (bottom panel) as a function of redshift is shown by the red lines. The differences between the two {\sc TinyTim} PSF models is well within requirements, even for the `S' galaxy. These results therefore confirm the conclusion of \citetalias{Semboloni13}  that the uncertainty in the HST PSF model has a negligible impact on the determination of the CG bias.

\begin{figure*}
  \hbox{
    \includegraphics[width=5.8cm]{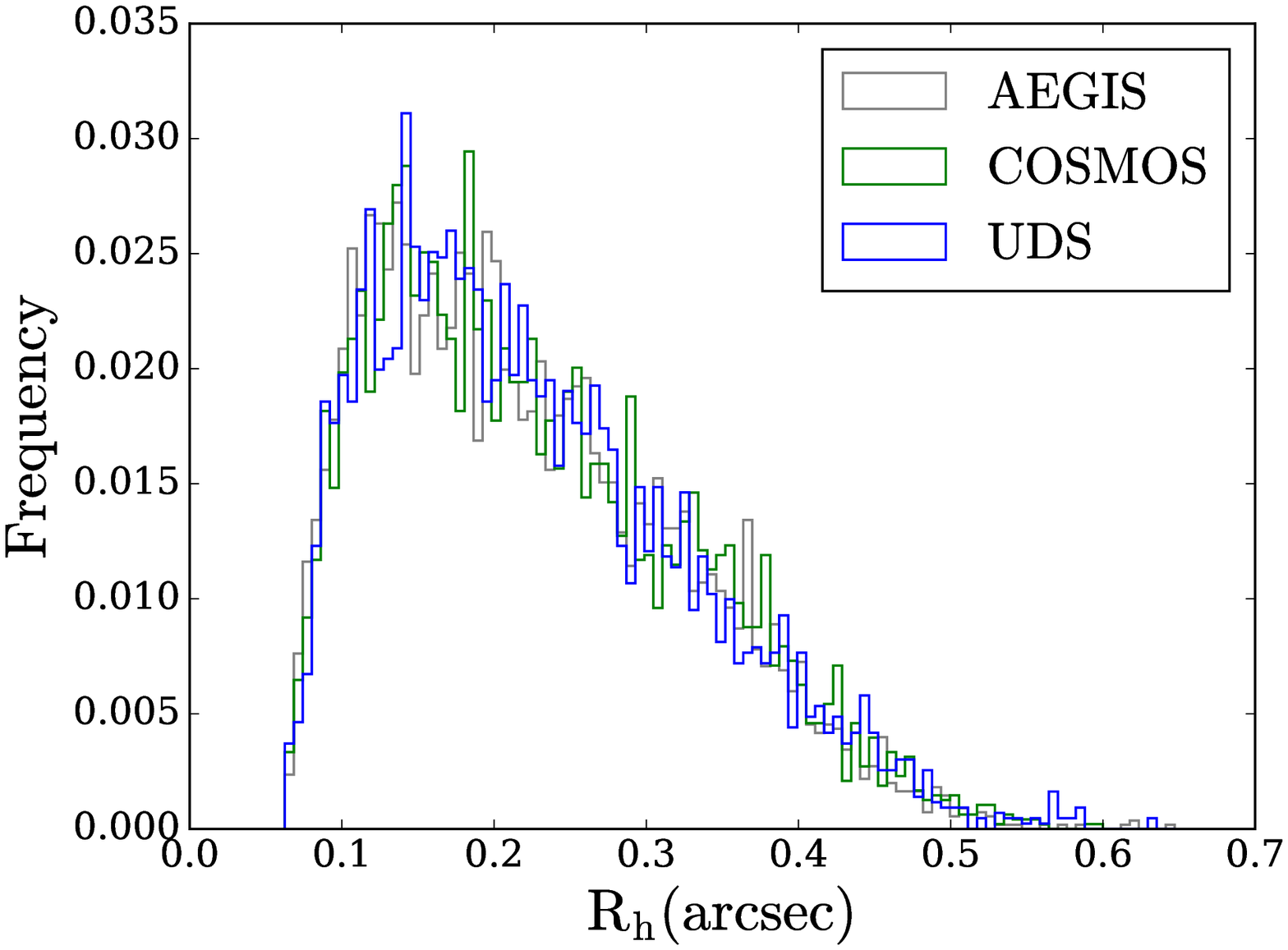}
    \includegraphics[width=5.8cm]{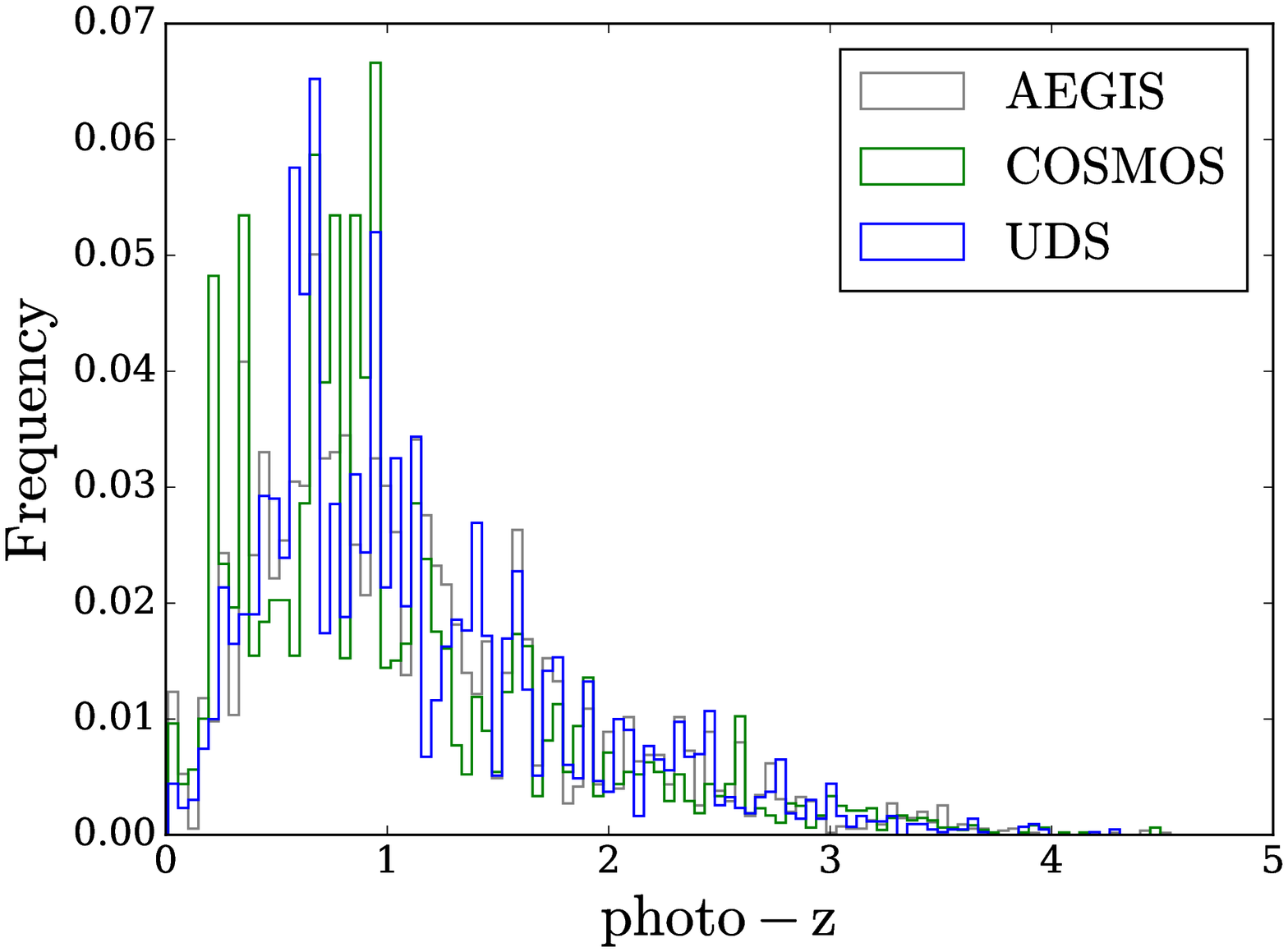}
    \includegraphics[width=5.8cm]{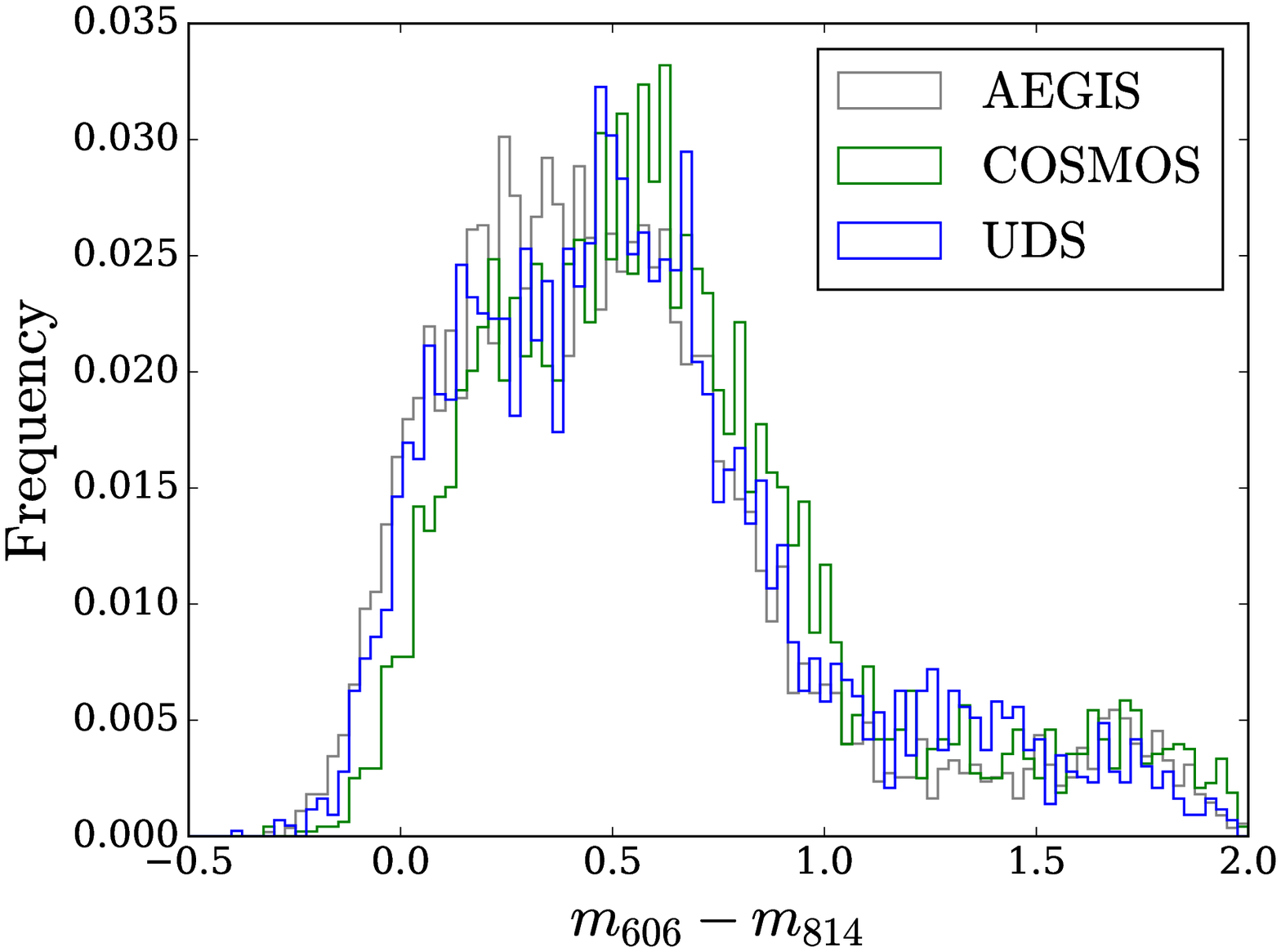}}
  \caption{Histogram of the distributions in observed half-light radii ($R_{\rm h}$; left panel),
  photometric redshift (middle panel) and $m_{606}-m_{814}$ colour (right panel) for the three
  CANDELS fields (AEGIS, COSMOS, UDS). We show results for galaxies with $m_{\rm VIS}<25$,
  where the results for the three fields are normalised by area.}
  \label{fig:datapro1}
\end{figure*}
\section{Measurement from HST observations}
\label{sec:candels}

In the previous section we confirmed the conclusion from
\citetalias{Semboloni13} that it is possible to determine the CG bias
from HST observations in the F606W and F814W filters. Importantly, we
demonstrated that the presence of noise in the actual data should not
bias the results significantly. We therefore proceed to determine the
expected CG bias in {\it Euclid} shape measurements using realistic
galaxy populations. To do so, we employ HST/ACS data taken in the
F606W and F814W filters in three of the CANDELS fields (AEGIS, COSMOS,
and UDS), which have a roughly homogeneous coverage in both bands
\citep[see][]{davis2007,grogin2011,Koekemoer11}.

We base our analysis on a tile-wise reduction of the ACS data, incorporating pointings that have at least four exposures to facilitate good cosmic ray removal, yielding combined exposure times of 1.3--2.3ks in F606W and 2.1--3.0ks in F814W.  We employ the updated correction for charge-transfer inefficiency from \cite{massey2014}, \textsc{MultiDrizzle} \citep{koekemoer2003} for the cosmic ray removal and stacking, as well as careful shift refinement, optimised weighting, and masking for stars and image artefacts as detailed in \cite{Schrabback10}. \cite{Schrabback16} created weak lensing catalogues based on these images, and we refer to this paper for more detail. \footnote{ Note that we do not apply the additional selection in measured half-light radius \mbox{$r_\mathrm{h}<7$} pixels employed in Schrabback et al. (2016) to not bias the overall sample compared to what may be used in a {\it Euclid} weak lensing analysis.}  We base our analysis on the galaxies that pass their source selection and apply additional magnitude cuts as detailed below. To investigate the dependence of the colour gradient influence on galaxy colour and redshift, we match this galaxy catalogue to the photometric redshift catalogue from \cite{skelton14}.

To resemble the selection of galaxies in the {\it Euclid} wide survey,
we estimate the flux in the VIS-band by linearly
interpolating the F606W and F814W fluxes from \cite{skelton14}
according to the effective wavelengths, where we adopted a central
wavelength of $735$\,nm for VIS.
We select galaxies brighter than $m_{\rm VIS}=25$.  The resulting sample sizes for the  three CANDELS fields are listed in Table~\ref{table:mag}. The number densities are in line with expectations for {\it Euclid} \citep{Laureijs11}.
Most galaxies in our sample are detected with an SNR$>15$, and we thus  expect to be able to determine the CG bias accurately.  In Fig.~\ref{fig:datapro1} we present histograms of some of the relevant galaxy properties for the three fields. We observe no significant differences, but note that we find more blue galaxies in AEGIS.

\begin{center}
\begin{table}
  \begin{tabular}{lllll}
    \hline
    Field               &AEGIS   &COSMOS   &UDS  &Total\\
    \hline
    Area [arcmin$^2$]  	 		&180      & 139     &146    & 465\\
    Total number                 		 	& 5518    & 4794  & 4311  &14623\\
    Number density [arcmin$^{-2}$]     & 30.7    & 34.5   & 29.5   &31.4\\
    \hline
  \end{tabular}
  \caption{\label{table:mag}
Properties of the sample of galaxies selected in the HST CANDELS fields. We
select galaxies with $m_{\rm VIS}<25$.}
\end{table}
\end{center}
\begin{figure}
  \includegraphics[width=\hsize]{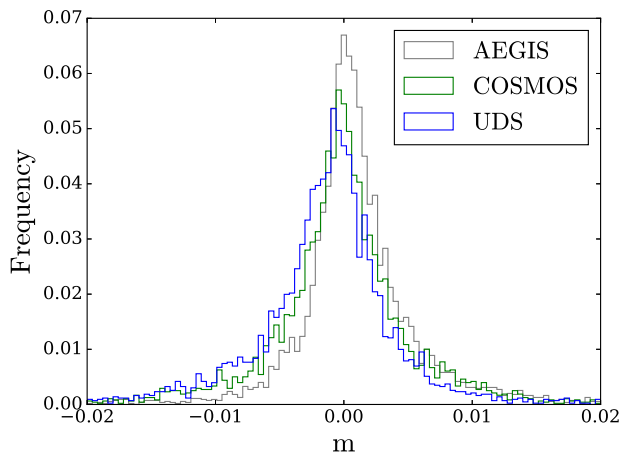}
\caption{Histogram of the estimated multiplicative bias caused by colour gradients using HST
observations. The results for the three different CANDELS fields are indicated by the different
colours.}
\label{fig:cgbhis}
\end{figure}

\subsection{CG bias from CANDELS}

We now proceed to apply the procedure we tested on synthetic galaxies to the HST observations to determine the expected CG bias for {\it Euclid}. We use the {\sc TinyTim} PSF when we fit the single component S{\'e}rsic models
to the observations using {\sc {\tt GALFIT}} (see Sect.~\ref{sec:psfmodel}).
We adopt priors on the S{\'e}rsic index ($0.5<n<5.0$), the effective radius ($1$ pixel$<r_e<50$ pixels) and axis ratio ($0.6<q<1.0$). As before, we approximate the {\it Euclid} PSF using Eqn.\,(\ref{eq:psfairy}).  As described in Sect.~\ref{sec:noisy} we interpolate the SED in each pixel of the model galaxy to generate a wavelength-dependent image, which is subsequently integrated and convolved to  create the images with and without colour gradients. We create images with six different orientations that are sheared to estimate the multiplicative shear bias $m$ caused by colour gradients.

Fig.~\ref{fig:cgbhis} shows the histogram of the CG bias for the three CANDELS fields that we study here.
Note that the observed distribution is slightly broadened due to noise in the HST images (cf. the red histogram in Fig.~\ref{fig:histogrambias}). The mean bias is $1.1\times 10^{-4}$ and the distribution is quite peaked, with biases less than 0.01 for 94\% of the galaxies.  The biases decrease by about a factor five when we double the width of the weight function that is used to measure the shapes. This demonstrates that the amplitude of the CG bias will be quite sensitive to the adopted weight function and thus depends on the adopted shape measurement algorithm: the CG bias will need to be determined for each algorithm that is applied to the data.

\begin{figure*}
\hbox{%
\includegraphics[width=0.5\hsize]{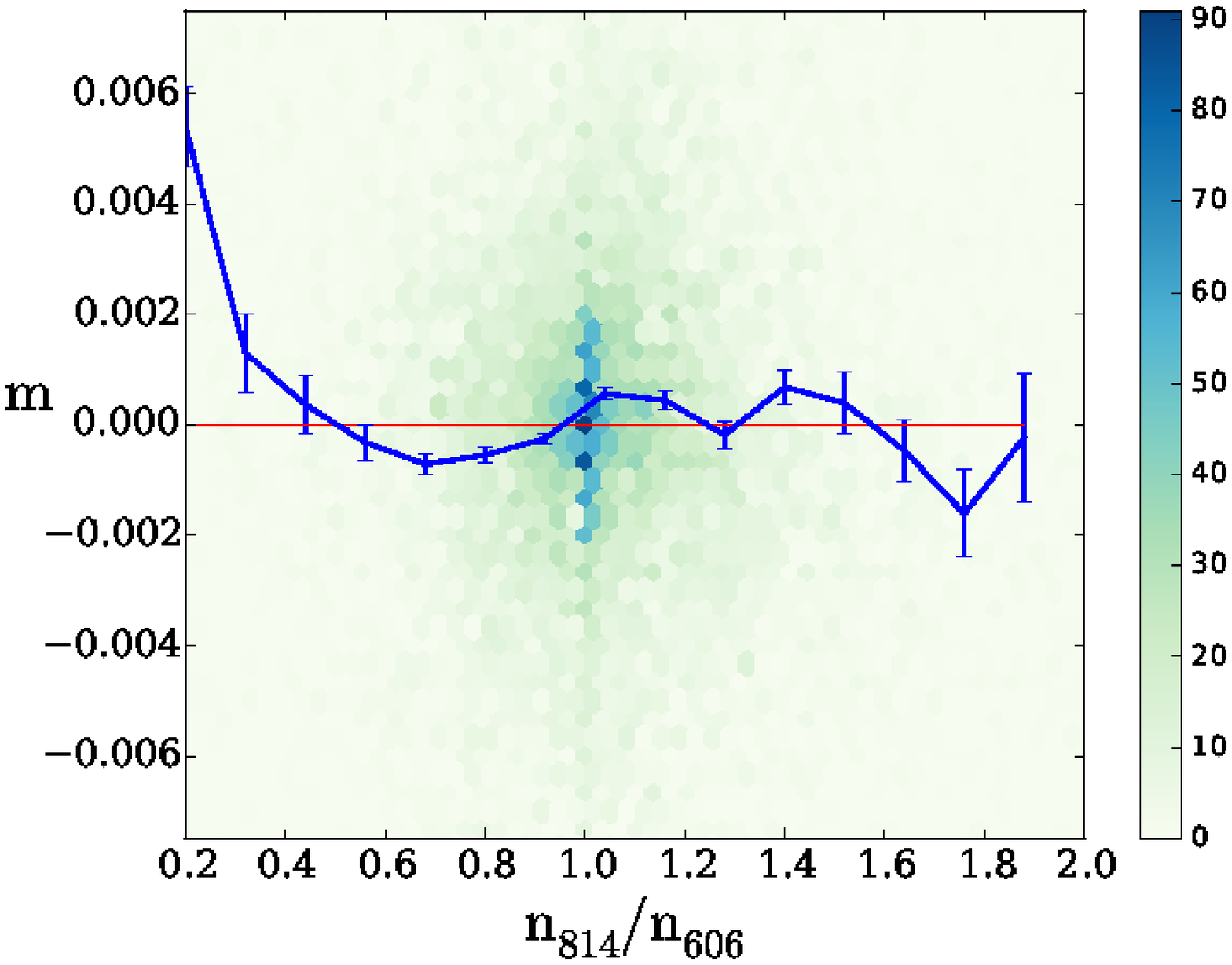}
\includegraphics[width=0.5\hsize]{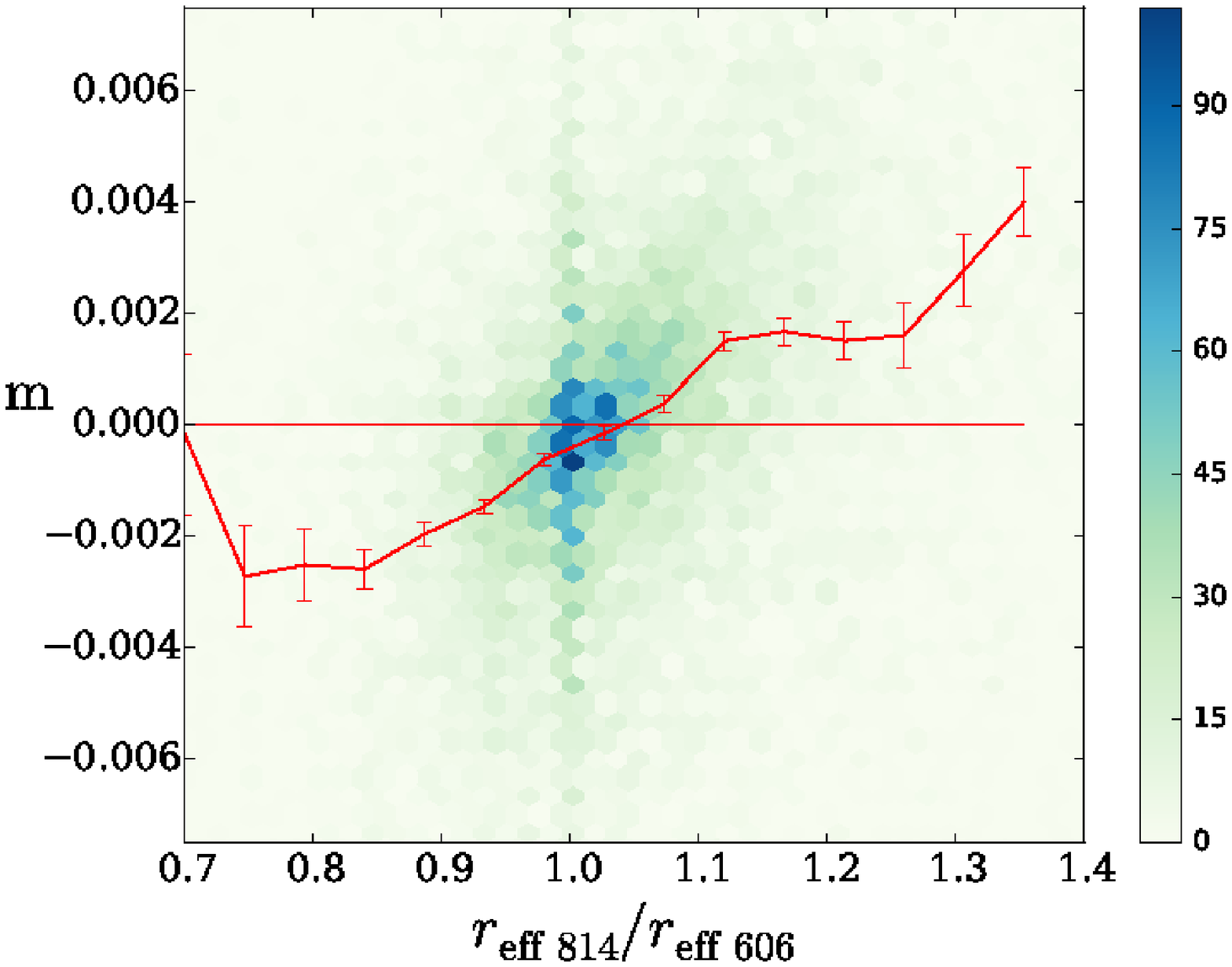}}
\caption{Multiplicative CG bias as a function of structural parameters
  in the fit to the surface brightness profiles in the F606W and F814W
  filters. {\it Top panel:} bias as a function of the ratio of the
  best fit S{\'e}rsic index in the F814W and F606W filters. The line
  with errorbars shows the average and its uncertainty. {\it Bottom
    panel:} bias as a function of effective radii in the F814W and
  F606W filters. We observe a clear trend in the average bias as a
  function of this ratio. The colour of the hexagon stands for the number of galaxies.}
\label{fig:cg2fitpar}
\end{figure*}

The amplitude of the CG bias depends on a number of parameters, such as the redshift and colour. Hence it is not sufficient to consider the average bias for the source sample, and we therefore explore such trends using our HST measurements. We first consider two quantities that should be directly related to
the CG bias, namely the ratio of the S{\'e}rsic index in the two HST filters and the ratio of the effective
radii in the two bands. The results are presented in Fig.~\ref{fig:cg2fitpar}. The top panel shows that the
average CG bias does not depend significantly on the ratio of S{\'e}rsic indices; we do observe a significant trend when we consider the ratio of effective radii (bottom panel). This is not surprising, because the bias in shape measurements depends to leading order on the galaxy size \citep{Massey13}. Note that the average CG bias in Fig.~\ref{fig:cg2fitpar} vanishes  when $r_{\rm eff,606}\approx r_{\rm eff, 814}$: in this case there should be no significant colour gradient (as the difference in S{\'e}rsic index has only a minor impact).

\begin{figure*}
\hbox{%
\includegraphics[width=0.5\hsize]{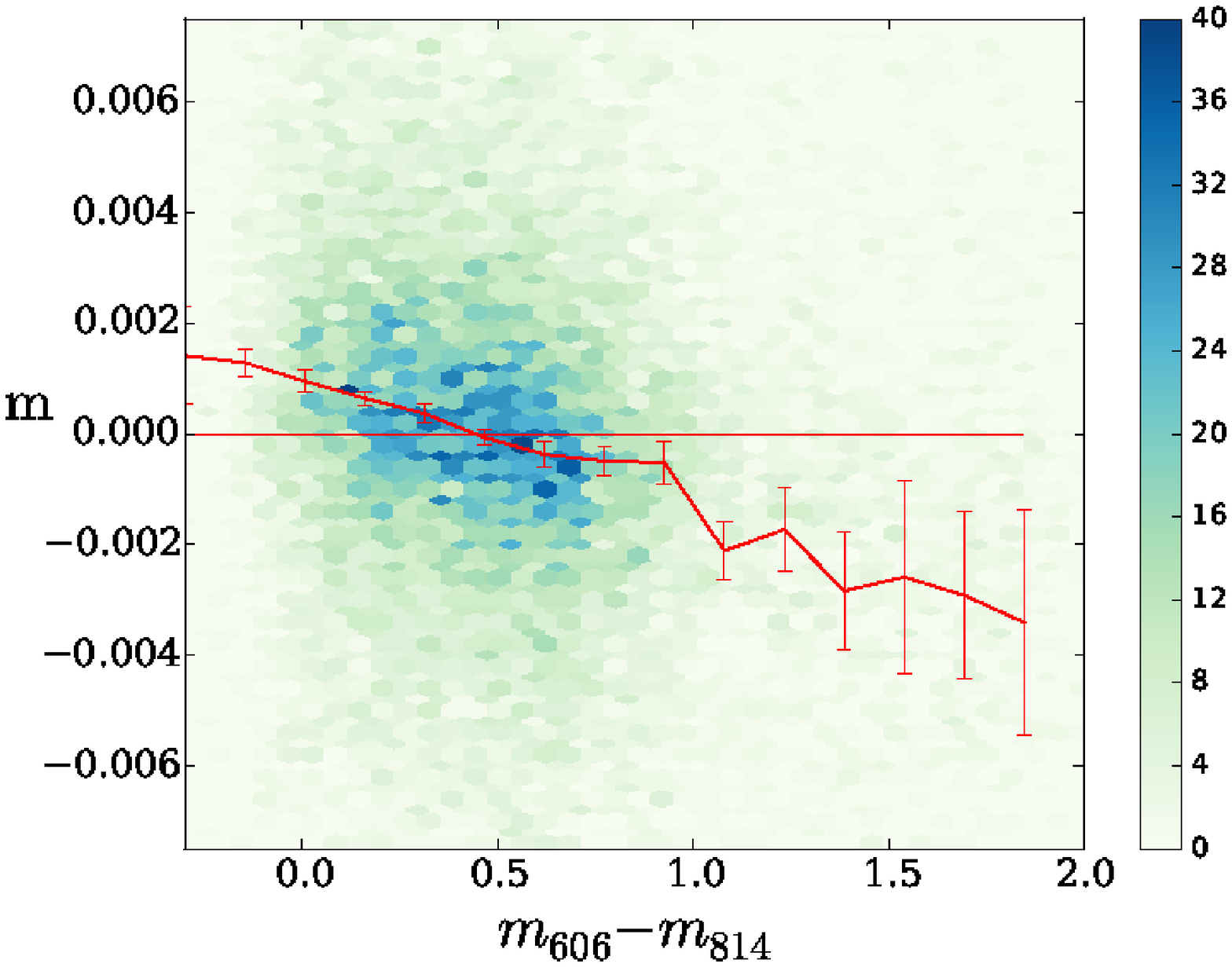}
\includegraphics[width=0.5\hsize]{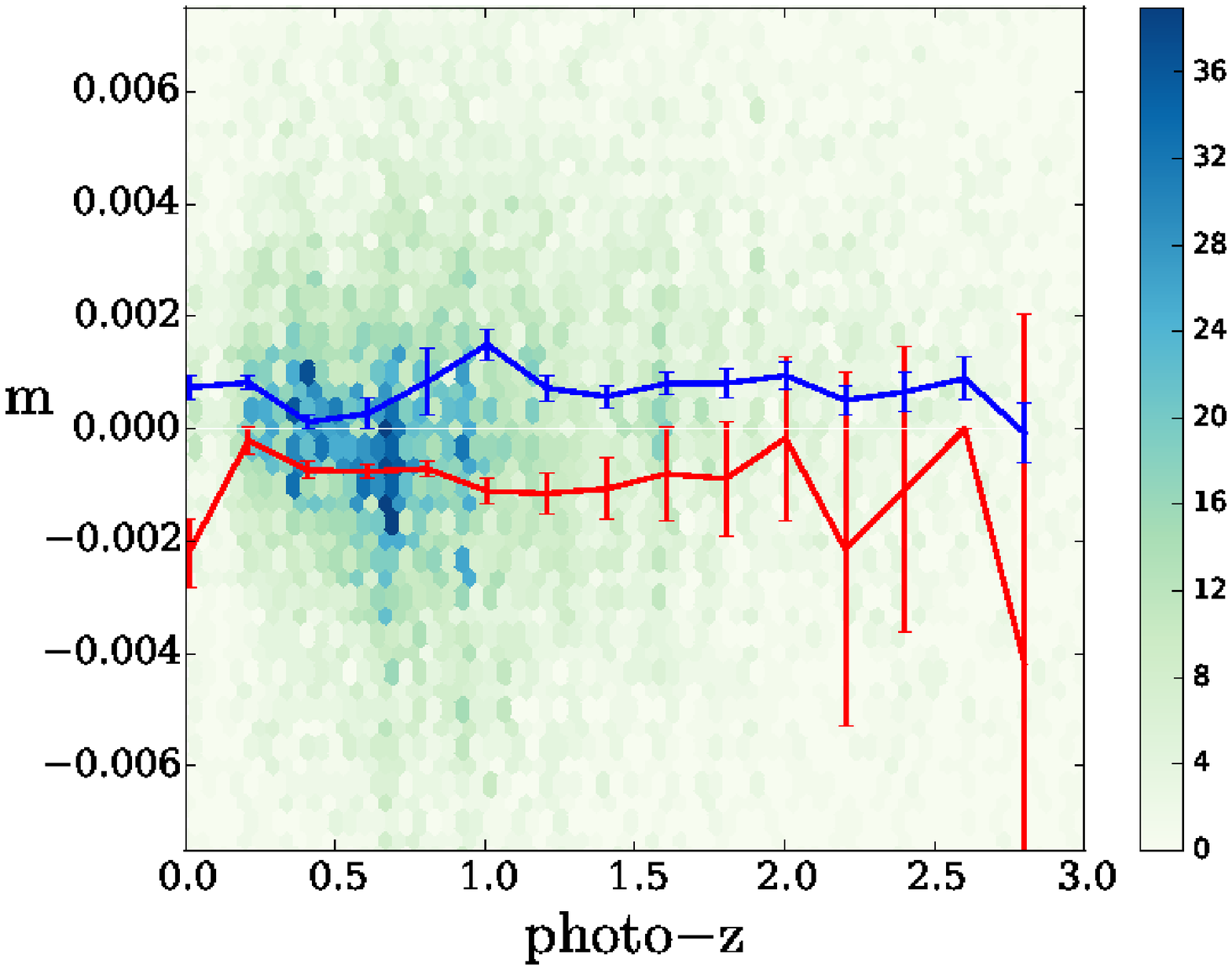}}
\caption{Multiplicative CG bias as a function of observed colour ($m_{606}-m_{814}$; left panel)
and redshift (right panel). We observe a clear trend of the average bias with colour (indicated by
the red points with error bars in the left panel). In the right panel we split the sample into red
($m_{606}-m_{814}>0.5$; red line) and blue ($m_{606}-m_{814}<0.5$; blue line) galaxies. The
variation with redshift is weak for both samples. The colour of hexagon stands for the number of galaxies.}
\label{fig:cg2color}
\end{figure*}

These structural parameters are, however, not observable using the {\it Euclid} data. Instead we proceed to examine trends with observable properties that correlate with the amplitude of the lensing signal, namely source redshift (the lensing signal is higher for more distant sources) and colour (as galaxies tend to be redder in high density regions). We show the CG bias as a function of the $m_{606}-m_{814}$ colour in the left panel of Fig.~\ref{fig:cg2color}, which shows that the average bias decreases for redder galaxies.

\begin{table}
\begin{center}
  \begin{tabular}{lccc}
    \hline
    photo-z    &Number &$\langle m\rangle$  &$\sigma_m$ \\
    \hline
    $0-0.4$    &399  &$-5.6\times10^{-4}$  &$0.0024$\\
    $0.4-0.8$  &3163 &$-7.1\times10^{-4}$  &$0.0036$\\
    $0.8-1.2$  &2960 &$-6.4\times10^{-4}$  &$0.0046$\\
    $>1.2$     &958  &$-1.4\times10^{-3}$  &$0.0062$\\
    \\
    $0-0.4$  &1513  &$6.2\times10^{-4}$  &$0.0026$\\
    $0.4-0.8$ &1154  &$1.9\times10^{-4}$  &$0.0029$\\
    $0.8-1.2$ &537  &$1.3\times10^{-3}$  &$0.0039$\\
    $>1.2$  & 3921  &$4.8\times10^{-4}$  &$0.0040$\\
    \hline
  \end{tabular}
  \caption{The number of objects, average and r.m.s. As the CG bias does not follow a Gaussian distribution, the value is estimated by the range that contains $68 \% $ of the measurements. CG bias in redshift bins for red (top half, $m_{606}-m_{814}>0.5$) and blue (bottom half, $m_{606}-m_{814}<0.5$) galaxies. }
  \label{table:calibration}
\end{center}
\end{table}

As the mean colour varies with redshift, we show the CG bias as a function of redshift in the right panel of
Fig.~\ref{fig:cg2color}.  Because the bias depends on colour, we split the sample into two groups. The average bias for the red galaxies ($m_{606}-m_{814}>0.5$) is indicated by the red line. The bias is negative and nearly constant over the redshift range of interest. Similar results are obtained for the blue galaxies
(defined as $m_{606}-m_{814}<0.5$), but in this case the mean bias is positive. The average CG biases and their dispersions for the two samples in various broad redshift bins are listed in Table~\ref{table:calibration}.

In additional tests, we also fit the galaxies with two component S{\'e}rsic models (similar to those used in Sect\,\ref{sec:noisy}), but this failed for a large fraction of galaxies ($\sim30\%$), since two S{\'e}rsic components may not be a good model for some, e.g. elliptical galaxies. The CG bias estimated from those galaxies fitted successfully shows a large scatter, but a similar relationship with colours of the galaxies as in the fitting with a single component. Moreover, as we test in the simulations, fitting the galaxies using one S{\'e}rsic component captures the main properties of the CG bias, we conclude that the results of fitting a single component can adequately represent the properties of the CG bias. However, more detailed studies are necessary for an accurate calibration of the bias.

\subsection{Use of morphological information}

\begin{figure*}
\hbox{%
\includegraphics[width=0.5\hsize]{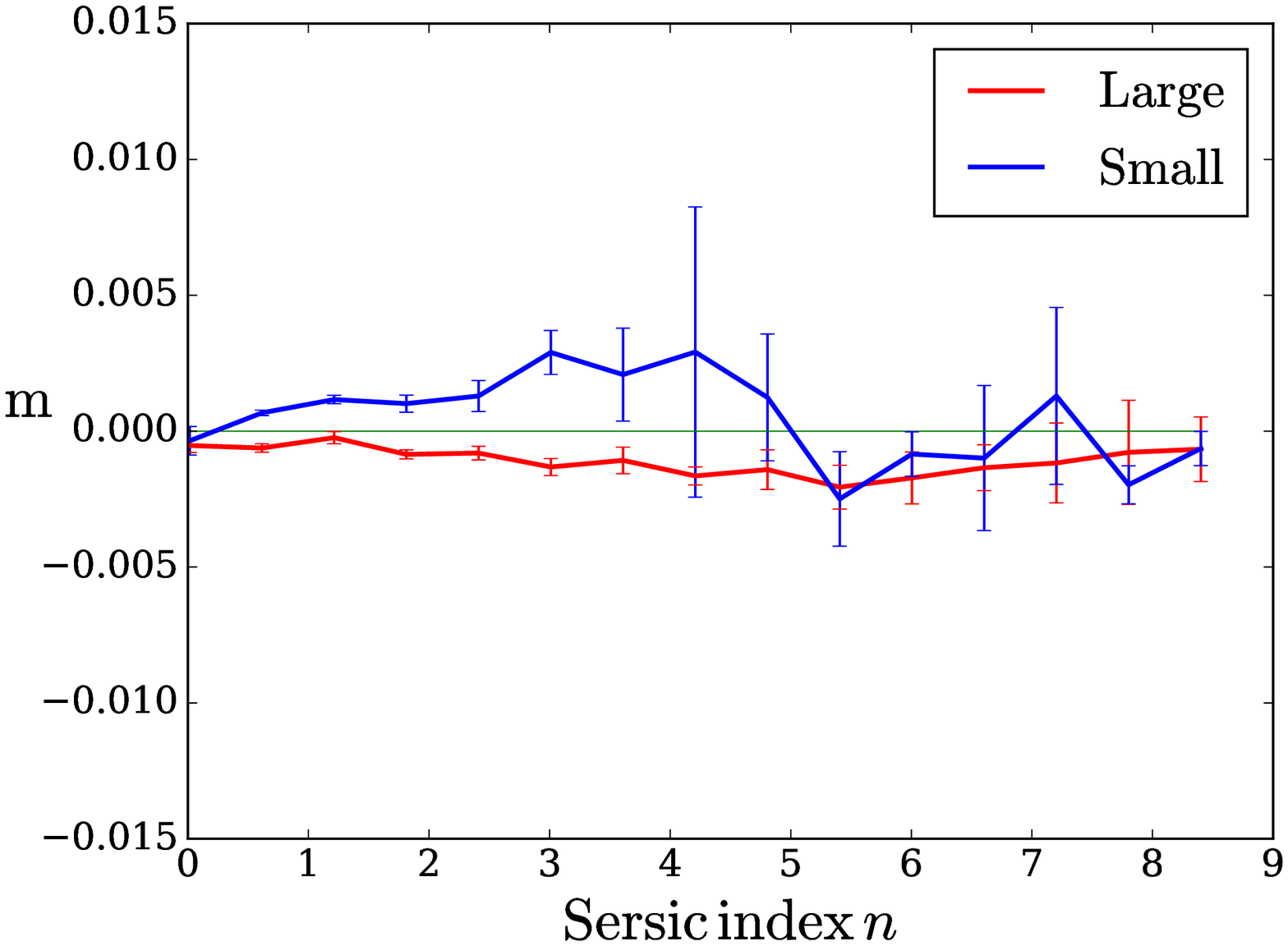}
\includegraphics[width=0.5\hsize]{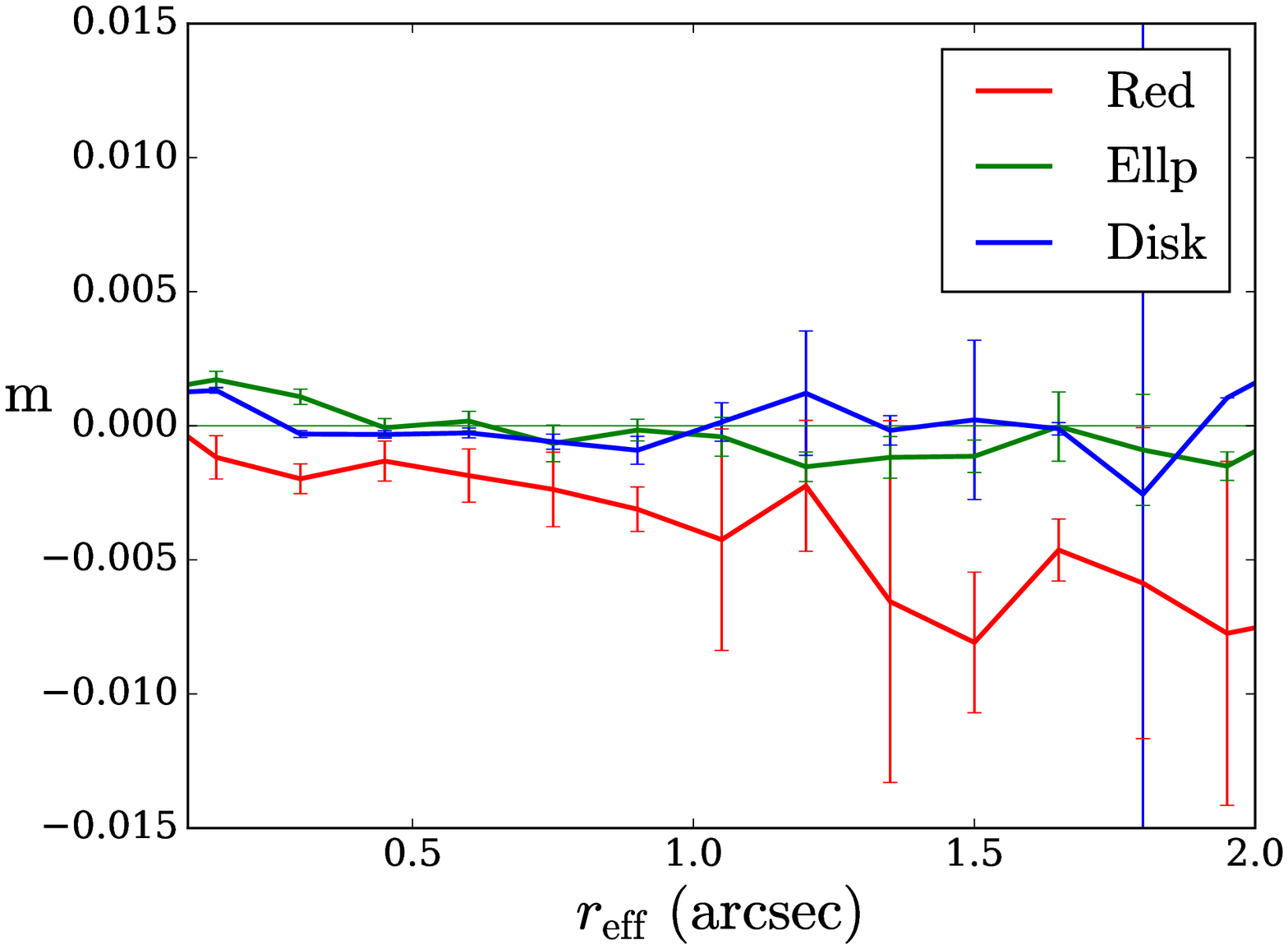}}
\caption{CG bias with S{\'e}rsic index (left) and effective radius (right)
  from the mock VIS images.
  In the top panel, the blue (red) is the average of small
  (big) galaxies. In the bottom panel, the red line is the average bias of
  red galaxies ($m_{606}-m_{814}>1$); the green line is that of elliptical
  galaxies ($n>2.75$); the blue line is for the disk galaxies.}
\label{fig:cg2re}
\end{figure*}

The lack of resolved multi-colour data from {\it Euclid} prevents us from
measuring colour gradients directly, but it may be possible to use some of the morphological
information that can be obtained from the VIS image. This is supported by the results
presented in Fig.~\ref{fig:cg2re}. The top panel shows the CG bias as a function of the
S{\'e}rsic index measured from the VIS images when we split the sample based on the
observed effective radii (galaxies with $r_{\rm eff}>0\farcs 35$ are classified as `large'
and the others as `small'). The large galaxies cover a large range of S{\'e}rsic index and have on average a negative average CG bias. Most of the small galaxies have small S{\'e}rsic indices ($<2.5$), and
the average bias is positive, with slightly larger values for $n\sim 4$.
In the bottom panel of Fig.~\ref{fig:cg2re} the galaxies are divided into three groups: red galaxies
with $m_{606}-m_{814}>1.0$; the remaining galaxies are subdivided into those with
large S{\'e}rsic indices ($n>2.25$, `early type galaxies') or small S{\'e}rsic indices (`disk galaxies').
The lines show the average CG bias as a function of effective radius.

These results suggest that the VIS image can provide additional information that can be
used in combination with the observed colour and redshift to refine the estimate of the
CG bias.  We find that the average bias is small for disk galaxies, as is the scatter
in the bias for small disk galaxies ($r_{\rm eff}<1''$). The early type galaxies cover
a large range in size, and the bias is significant for the reddest galaxies, albeit with
increased scatter. Further trends could be explored when larger multi-colour HST data sets
are considered. In particular machine-learning techniques could be used to explore
parameter combinations that reduce the scatter in the estimate for the CG bias for individual galaxies.

We can use the values for the scatter in the CG bias (listed in Table~\ref{table:calibration}) to
estimate the number of galaxies that we need to calibrate the bias with high precision. We estimate
we need approximately $400$ galaxies for each type of galaxy in every redshift bin. If we consider
relatively wide bins, for instance, two types of colour: red and blue; five different sizes from
about $0.1$ arcsec to $1.0$ arcsec (Fig.\,\ref{fig:cg2re}), and five redshift bins, we require at least
$40\,000$ galaxies. The numbers increase if we wish to use a finer SED classification (rather than
simply blue and red). In our study we restricted the observations to three of the CANDELS fields
with homogeneous coverage and included only the area with high-quality redshift estimates from
3D-HST, yielding a total sample of approximately $15\,000$ galaxies. When improved data for photometric
redshifts are obtained for the parts of these fields outside 3D-HST, and when the additional
two CANDELS fields are included, the total galaxy sample approximately matches the required number
(see also Table~4 in \citetalias{Semboloni13}, where we exclude the F850LP observations given the
significantly lower SNR). Hence, we conclude that a coarse correction for CG bias can be inferred
from these data. However, a larger number of galaxies is needed for the CG calibration if a finer
SED classification (rather than simply blue and red) or a larger number of redshift bins is
used. Such a finer calibration would be enabled by additional HST observations. These must target
representative ``blank fields", include HST coverage in bands that fully cover the VIS filter,
and have high quality redshift estimates available.

\section{Summary and Discussion}

The next generation of wide area deep imaging surveys will dramatically improve the precision with which the correlations in galaxy shapes caused by weak gravitational lensing will be measured. However, to exploit these data, it is paramount that instrumental effects are accounted for. Many of these could hitherto be ignored, but this will not be the case any longer in the case of {\it Euclid}, {\it WFIRST} and LSST.
Although the shape measurements greatly benefit from the compact diffraction-limited point spread function (PSF) in space-based observations, it is important that chromatic effects are accounted for. This is particularly relevant in the case of {\it Euclid}, which employs a broad pass band to maximise the signal-to-noise ratio of the measurements. This enhances its sensitivity to spatial variations in the colours of galaxies,
which result in biases in the inferred lensing signal, unless accounted for.

In this paper we showed that the CG bias can be quantified with high accuracy using available multi-colour {\it Hubble} Space Telescope (HST) data. We validated our approach against earlier work presented by
\citetalias{Semboloni13}. Our implementation is different but yields consistent results (note that our definition does have the opposite sign compared to \citetalias{Semboloni13}). We also extended the analysis to higher order lensing effects, which occur in high-density regions. Flexion leads to enhanced CG bias, but only close to the lens. Hence this can be relevant for small-scale galaxy-galaxy lensing studies with {\it Euclid}. It can, however, be safely ignored in the case of cluster studies and cosmic shear.

Previous studies ignored the potential detrimental effect of noise in the HST observations that are used to infer the CG bias. Fortunately, our results indicate that this does not change the CG bias estimates significantly for {\it Euclid} source galaxies, given the noise levels in the HST data used. It does slightly increase the noise in the shape measurements, but the biases for individual galaxies are generally well below 1\%.
The inferred bias depends strongly on the weight function used to measure shapes. Consequently the CG bias will need to be determined for each shape measurement algorithm separately.

After testing our approach on simulated data, we measured the CG bias using HST/ACS observations
in the F606W and F814W passbands. We used observations from three CANDELS fields, which have
fairly uniform coverage in both filters, and for which redshift information is available. This allowed us to quantify the CG bias as a function of redshift and colour. As expected, the CG bias correlates with observed colour, but the dependence with redshift is weak. Although the observed biases are small, they cannot be ignored for {\it Euclid}. Although further study is required, we find that it should be possible to reduce the bias for individual galaxies by using morphological information (e.g. S{\'e}rsic index, effective radius) that can be obtained from the {\it Euclid} data themselves.

We use the observed trends and scatter in the bias to estimate the number of galaxies for which similar high-quality HST data are needed. This leads to a minimum requirement of more than $40\,000$ galaxies for a coarse correction. HST has covered sufficient area in the CANDELS fields in F606W and
F814W to approximately match this number, but not yet all of this area is covered by sufficient
multi-wavelength data for high quality redshift estimates (especially outside of the HST/WFC3 footprints). Additional HST observations would provide an improved CG calibration by enabling a
finer binning in galaxy redshift and SED. This would be achieved most effectively by complementing
fields that are already covered by one of the required HST filters and for which high-quality redshift information is available; for instance by adding F606W (or F625W) observations to the wider ACS/F814W
mosaic in the HST COSMOS survey \citep{scoville07hst}.


\section*{Acknowledgments}

We like to thank the referee for very useful comments to improve our draft.
We would like to thank Emiliano Merlin, Marco Castellano for help with
{\sc SExtractor} and {\sc {\tt GALFIT}}, Lance Miller, Thomas
Kitching, Gary Berstein, Adam Rogers, Ignacio Ferreras and also the
members of the {\it Euclid} Consortium for useful discussions. XE and VFC are
funded by Italian Space Agency (ASI) through contract Euclid -IC
(I/031/10/0) and acknowledge financial contribution from the agreement
ASI/INAF/I/023/12/0. XE is also partly support by NSFC Grant
No. 11473032. HH acknowledges support from the European Research
Council FP7 grant number 279396. TS acknowledges support from the
German Federal Ministry for Economic Affairs and Energy (BMWi)
provided via DLR under project no. 50QE1103. MV acknowledges support from Funda??o para a Ci¨ºncia
 e a Tecnologia (FCT) through the research grant IF/01518/2014. JR is supported by JPL,
which is run by Caltech under a contract for NASA, and is supported by
grant NASA ROSES 12-EUCLID12- 0004. The fast Fourier transforms are
supplied by the FFTW library \citep{fftw05}. We use CFITSIO
\citep{1999ASPC..172..487P} for the FITS file.

\bibliographystyle{aa}
\bibliography{cbias}

\end{document}